\documentclass[a4paper,12pt]{article}
\usepackage{amsmath}
\usepackage{amsfonts}
\usepackage{times}
\usepackage{amsthm}
\usepackage{epsfig}
\newcommand{\DATUM}{November 16, 2008}              
\newcommand{\comma}{\: ,}     
\newcommand{\period}{\: .}    
\newcommand{\Proof}{\noindent\emph{Proof. }}              
\newcommand{\QED}{\hspace*{\fill}\mbox{$\Box$}}           
\newcommand{\eps}{{\varepsilon}}
\newcommand{\vphi}{{\varphi}}           
\newcommand{\Om}{\Omega}                
\newcommand{\om}{\omega}
\newcommand{\la}{\langle}
\newcommand{\ra}{\rangle}
\newcommand{\C}{C_\chi}
\newcommand{\one}{\mathbf{1}}

\newcommand{\cB}{\mathcal{B}}
\newcommand{\cD}{\mathcal{D}}

\newcommand{\cF}{\mathcal{F}}

\newcommand{\cH}{\mathcal{H}}

\newcommand{\cM}{\mathcal{M}}         
\newcommand{\cO}{\mathcal{O}}         

\newcommand{\cR}{\mathcal{R}}
\newcommand{\cS}{\mathcal{S}}
\newcommand{\cT}{\mathcal{T}}
\newcommand{\cU}{\mathcal{U}}

\newcommand{\cW}{\mathcal{W}}

\newcommand{\field}[1]{\mathbb{#1}}

\newcommand{\RR}{\field{R}}     
\newcommand{\NN}{\field{N}}     
\newcommand{\CC}{\field{C}}     
\newcommand{\fh}{\mathfrak{h}}  
\newcommand{\bchi}{{\overline{\chi}}}

\newcommand{\bpi}{\bar{\pi}}
\newcommand{\uw}{{\underline w}}
\newcommand{\umpnq}{{\underline{m,p,n,q}}}
\newcommand{\upq}{{\underline{0,p,0,q}}}
\newcommand{\tuw}{{\underline{\tilde{w}}}}
\newcommand{\huw}{{\underline{\hat{w}}}}
\newcommand{\hw}{\hat{w}}

\newcommand{\tV}{\widetilde{V}} 
\newcommand{\tW}{\widetilde{W}} 
\newcommand{\tk}{\tilde{k}}

\newcommand{\tr}{\tilde{r}}
\newcommand{\tw}{\widetilde{w}}
\newcommand{\tx}{\tilde{x}}

\newcommand{\tpi}{\tilde{\pi}}

\newcommand{\sym}{\mathrm{sym}}
\newcommand{\const}{\mathrm{const}}
\newcommand{\red}{\mathrm{red}}
\newcommand{\rIm}{\mathrm{Im}}
\newcommand{\rRe}{\mathrm{Re}}               
\newcommand{\Ran}{\mathrm{Ran}}              

\newcommand{\cirS}{\mathop{\bigcirc\kern -.73em {\scriptstyle{\rm S}}}}

\newcommand{\dist}{{\rm dist}}

\newcommand{\dom}{\mathrm{dom}}
\newcommand{\supp}{\mathrm{supp}}

\newcommand{\cern}{\mathrm{Ker}}
\newcommand{\op}{\mathrm{op}}
\newcommand{\at}{{p}}

\newcommand{\hf}{H_f}

\newcommand{\lb}{\left(}
\newcommand{\rb}{\right)}
\renewcommand{\thesection}
{\Roman{section}}                      
\renewcommand{\theequation}
{\thesection.\arabic{equation}}        
\newcommand{\secct}[1]{\section{#1}
\setcounter{equation}{0}}              

\newtheorem{theorem}{Theorem}[section]         
\newtheorem{lemma}[theorem]{Lemma}             
\newtheorem{corollary}[theorem]{Corollary}     
\newtheorem{remark}[theorem]{Remark}           
\newtheorem{proposition}[theorem]{Proposition} 
\theoremstyle{plain}
\begin{document}
\bibliographystyle{plain}
\setcounter{page}{0}
\thispagestyle{empty}

\title{Ground State and
Resonances in the Standard Model of the Non-relativistic QED}
\author{Israel Michael Sigal
\thanks{Supported by NSERC Grant No. NA7901}
\thanks{webpage: www.math.toronto.edu/sigal}
\thanks{Visiting IAS, Princeton, NJ, U.S.A.}\\
\small{Department ~of Mathematics, Univiversity of Toronto, Toronto,
Canada}\\[-1ex]}

\date{\DATUM}

\maketitle

  \centerline{\textit{To J\"urg and Tom, with admiration}}

\begin{abstract}
  We prove existence of a ground state and resonances in the standard model of the
  non-relativistic quantum electro-dynamics (QED).
  To this end we introduce a new canonical transformation of QED Hamiltonians and use
  the spectral renormalization group technique with a new choice of Banach spaces.

\end{abstract}
%

\setcounter{page}{1}

\secct{Introduction} \label{sec-I}
\textbf{Problem and outline of the results}. Non-relativistic
quantum electro-dynamics (QED) describes the processes of
emission and absorption of radiation by systems of matter, such as
atoms and molecules, as well as other processes arising from
interaction of the quantized electro-magnetic field with
non-relativistic matter.
The mathematical framework of this theory  is well established. It
is given in terms of the time-dependent Schr\"odinger equation,
$$i\partial_t\psi= H^{SM}_g\psi,$$ where $\psi$ is a differentiable path
in the Hilbert space $\cH=\cH_{p}\otimes\cH_{f}$, which is the
tensor product of the state spaces of the matter system $\cH_{p}$,
and the quantized electromagnetic field $\cH_{f}$ and $H^{SM}_g$ is
the standard quantum Hamiltonian given by\footnote{For discussion of
physics emerging out of this Hamiltonian see
\cite{Cohen-TannoudjiDupont-RocGrynberg1991,
Cohen-TannoudjiDupont-RocGrynberg1992}. }
\begin{equation}\label{Hsm}
H^{SM}_g=\sum\limits_{j=1}^n{1\over 2m_j}
(i\nabla_{x_j}+gA(x_j))^2+V(x)+H_f.
\end{equation}
It acts on the Hilbert space $\cH=\cH_{p}\otimes\cH_{f}$. 
Here 
the superindex $SM$ stands for 'standard model' and, as explained
below, $g$ is related to the particle
charge, or, more precisely, to the fine-structure constant. 
The remaining symbols are defined below. (To simplify the exposition
we omitted the interaction of the spin with magnetic field -
$\sum\limits_{j=1}^n \frac{g}{2m_j}\sigma_j \cdot \textrm{curl}
A(x_j)$. It can be easily incorporated into our analysis.)

This model has been extensively studied in the last decade, see the
book \cite{Spohn} and reviews \cite{ Arai1999, Hiroshima5,
Hiroshima7, Griesemer}  and references therein for a list of early
contributions.

For a large class of potentials $V(x)$, including Coulomb
potentials, and under an ultra-violet cut-off, the operator
$H^{SM}_g$ is self-adjoint (see e.g. \cite{BachFroehlichSigal1999,
Hiroshima4}). The stability of the system under consideration is
equivalent to the statement of existence of the ground state of
$H^{SM}_g$, i.e. an eigenfuction with the smallest possible energy.
The physical phenomenon of radiation is expressed mathematically as
emergence of resonances out of
excited states of a particle system
due to coupling of this system to the quantum electro-magnetic
field.
We define the resonances and discuss their properties below.



In this paper we prove existence of the ground state and resonance
states of $H^{SM}_g$ originating from the ground state and from
excited states of the particle system.
Our approach provides also an effective way to compute the ground
states and resonance states and their eigenvalues. We do not impose
any extra conditions on $H^{SM}_g$, except for smallness of the
coupling constant $g$ and an ultraviolet cut-off in the interaction.

The existence (and uniqueness) of the ground state was proven by
compactness techniques in \cite{GriesemerLiebLoss, LiebLoss, BachFroehlichSigal1999, Hiroshima1,
Hiroshima2, Hiroshima3, HiroshimaSpohn,  AraiHirokawa} and in a constructive way, in
\cite{BachFroehlichPizzo1}\footnote{Analyticity of the ground state eigenvalues in parameters was proven in \cite{GriesemerHasler2}. }. The existence of the resonances was
proven so far only for confined potentials (see
\cite{BachFroehlichSigal1998a,BachFroehlichSigal1998b} and, for a
book exposition, \cite{GustafsonSigal}). (Note that the papers
\cite{BachFroehlichSigal1999, Hiroshima1, GriesemerLiebLoss,
LiebLoss, HiroshimaSpohn} include the interaction of the spin with
magnetic field 
in the Hamiltonian, while the present paper omits it.)

%
%

Our proof contains two new ingredients: a new canonical
transformation of the Hamiltonian $H^{SM}_g$ (which we call the
generalized Pauli-Fierz transformation, Section \ref{sec-II}) and
new -- momentum anisotropic -- Banach spaces for the spectral
renormalization group (RG) which allow us to control the RG flow for
more singular coupling functions. 
A part of this paper which deals with adapting and clarifying some
points of the RG technique for the present situation (see Appendix
B) is rather technical but can be used in other problems of
non-relativistic QED.

\textbf{Standard model}. We now describe the standard model of
non-relativistic QED. 
We use the units in which the Planck constant divided by $2\pi$, the
speed of light and the electron mass are equal to $1(\ \hbar=1$,
$c=1$  and $m=1$). In these units the electron charge is equal to
$-\sqrt{\alpha}$,
where $\alpha =\frac{e^2}{4\pi \hbar c}\approx {1\over 137}$, the
fine-structure constant, and the distance, time and energy are
measured in units of $\hbar/mc =3.86 \cdot 10^{-11}cm,\ \hbar/mc^2
=1.29 \cdot 10^{-21} sec$ and $mc^2 = 0.511 MeV$, respectively
(natural units).

We consider the matter system consisting of $n$ charged particles
interacting with quantized electromagnetic field. The particles have
masses $m_j$
and positions $x_j$, where $j=1, ..., n$. We write
$x=(x_1,\dots,x_n)$. The total potential of the particle system is
denoted by $V(x)$. The Hamiltonian operator of the particle system
alone is given by
\begin{equation} \label{Hp}
H_p:=-\sum\limits_{j=1}^n {1\over 2m_j} \Delta_{x_j}+V(x),
\end{equation}
where $\Delta_{x_j}$ is the Laplacian in the variable $x_j$.  This
operator acts on a Hilbert space of the particle system, denoted by
$\cH_{p}$, which is either $L^2(\mathbb{R}^{3n})$ or a subspace of
this space determined by a symmetry group of the particle system. We
assume that $V(x)$ is real and s.t. the operator $H_p$ is
self-adjoint on the domain of $\sum\limits_{j=1}^n {1\over 2m_j}
\Delta_{x_j}$.

The
quantized electromagnetic field
is described by the \textit{quantized vector potential}
\begin{equation}\label{A}
A(y)=\int(e^{iky}a(k)+e^{-iky}a^*(k))\chi(k){d^3k\over \sqrt{|k|}},
\end{equation}
in the Coulomb gauge ($\textrm{div} A(y)=0$). Here $\chi$ is an
\textit{ultraviolet cut-off}: $\chi(k)={1\over (2\pi)^3 \sqrt{2}}$
in a neighborhood of $k=0$ and it vanishes sufficiently fast at
infinity (we comment of this below).
The dynamics of the quantized electromagnetic field is given through
the quantum Hamiltonian
\begin{equation} \label{Hf}
\hf \ = \ \int d^3 k \; \om(k) a^*(k) \cdot  a(k) ,
\end{equation}
where $\om(k) \ = \ |k|$ is the dispersion law connecting the energy
of the field quantum with its wave vector $k$. Both, $A(y)$ and
$H_f$, act on the Fock space $\cH_{f}\equiv \cF$. Thus the Hilbert
space of the total system is ${\mathcal H}:= {\mathcal H}_p \otimes
{\mathcal F}$.

Above, $a^*(k)$ and $a(k)$ denote the creation and annihilation
operators on $\cF$. The families $a^*(k)$ and $a(k)$ are
operator-valued generalized, transverse vector fields:
$$a^\#(k):= \sum_{\lambda \in \{-1, 1\}}
e_{\lambda}(k) a^\#_{\lambda}(k),$$ where $e_{\lambda}(k)$ are
polarization vectors, i.e. orthonormal vectors in $\mathbb{R}^3$
satisfying $k \cdot e_{\lambda}(k) =0$, and $a^\#_{\lambda}(k)$ are
scalar creation and annihilation operators satisfying canonical
commutation relations. The right side of \eqref{Hf} can be
understood as a weak integral. See Supplement A for a brief review
of definitions of the Fock space, the creation and annihilation
operators
and the operator $\hf$.

The Hamiltonian of the total system, matter and the radiation field,
is 
given by \eqref{Hsm}. 
First, we consider \eqref{Hsm} for an atom or molecule. Then, in the
natural units,  $g= \sqrt{\alpha}$ and $V(x)$, which is the total
Coulomb potential of the particle system, is proportional to
$\alpha$. Rescaling $x \rightarrow \alpha^{-1} x$ and $k \rightarrow
\alpha^2 k$ we arrive at \eqref{Hsm} with $g:= \alpha^{3/2}$, $V(x)$
of the order $O(1)$\footnote{In the case of a molecule in the
Born-Oppenheimer approximation, the resulting $V(x)$ also depends on
the rescaled coordinates of the nuclei. } and $A(x)$ replaced by
$A'(x)$, where $A'(x) = A(\alpha x)|_{\chi(k) \rightarrow \chi'(
k)}$, where $\chi'( k):=\chi(\alpha^2 k)$ (see
\cite{BachFroehlichSigal1995,
BachFroehlichSigal1999}). 
%
After that we drop the prime in the vector potential $A'(x)$ and the
ultraviolet cut-off $\chi'(x)$ (see a discussion of the latter below). 
Finally, we relax the restriction on $V(x)$
by considering the standard generalized $n$-body potentials (see
e.g. \cite{HunzikerSigal}):
\begin{itemize}
\item [(V)] $V(x) = \sum_i W_i(\pi_i x)$, where $\pi_i$ are a linear maps from
$\mathbb{R}^{3n}$ to $\mathbb{R}^{m_i},\ m_i \le 3n $ and $ W_i$ are
Kato-Rellich potentials (i.e. $W_i(\pi_i x) \in
L^{p_i}(\mathbb{R}^{m_i}) + (L^\infty (\mathbb{R}^{3n}))_\eps$ with
$p_i=2$ for $m_i \le 3,\ p_i>2 $ for $m_i =4$ and $p_i \ge m_i/2$
for $m_i > 4$, see \cite{RSIV, HislopSigal}).
\end{itemize}

Under the assumption (V), the operator $H^{SM}_g$ is self-adjoint.
In order to tackle the resonances we choose the ultraviolet cut-off,
$\chi(k)$, so that
\begin{itemize}
\item [] The function $\theta \rightarrow \chi(e^{-\theta} k)$  has an
analytic continuation from the real axis, $\mathbb{R}$, to the strip
$\{\theta \in \mathbb{C} | |\rIm\ \theta | < \pi/4 \}$ as a $L^2
\bigcap L^\infty (\mathbb{R}^3)$ function, \end{itemize} e.g.
$\chi(k)= e^{-|k|^2/K^2}$, and we assume that the potential, $V(x)$,
satisfies the condition:
\begin{itemize}
\item [(DA)] The
the particle potential $V(x)$ is dilation analytic in the sense that
the operator-function
$\theta \rightarrow V(e^{\theta} x) (-\Delta +1)^{-1}$ has an
analytic continuation from the real axis, $\mathbb{R}$, to the strip
$\{\theta \in \mathbb{C} | |\rIm\ \theta | < \theta_0 \}$ for some
$\theta_0 > 0$.
\end{itemize}

In order not to deal with the problem of center-of-mass motion,
which is not essential in the present context, we assume that either
some of the particles (nuclei) are infinitely heavy or the system is
placed in a binding, external potential field. This means that the
operator $H_p$ has isolated eigenvalues below its essential
spectrum. However, the techniques developed in this paper can be
extended to translationally invariant particle systems (see
\cite{Faupin2007}).

\textbf{Ultra-violet cut-off}. Finally, we comment on the
ultra-violet cut-off $\chi(k)$ introduced in \eqref{A}. This cut-off
is introduced for the model to be well-defined. Assuming $\chi$
decays on the scale $k_c$, in order to correctly describe the
phenomena of interest,  such as emission and absorption of
electromagnetic radiation,  i.e. for optical and rf modes, we have
to assume that the cut-off energy, $\hbar c k_c,\ $ is much greater
than the characteristic energies of the particle motion (we
reintroduced the Planck constant, $\hbar$, and speed of light, $c$,
for a moment). The latter motion takes place on the energy scale of
the order of the ionization energy, i.e. of the order $\alpha^2
mc^2$. Thus we have to assume $\alpha^2 mc^2 \ll \hbar c k_c$.

On the other hand, for energies higher than the rest energy of the
the electron ($mc^2$) the relativistic effects, such as
electron-positron pair creation, vacuum polarization and
relativistic recoil, take place. Thus it makes sense to assume that
$\hbar c k_c \ll mc^2$. Combining the last two conditions we arrive
at the restriction  $\alpha^2 mc^2 \ll \hbar c k_c \ll mc^2$ or
    $\alpha^2 mc/\hbar \ll  k_c \ll   mc/\hbar.$
In our units this reads  $$\alpha^2  \ll  k_c \ll 1.$$ After the
rescaling $x \rightarrow \alpha^{-1} x$ and $k \rightarrow \alpha^2
k$ performed above the new cut-off momentum scale, $k'_c=
\alpha^{-2}k_c$, satisfies
$$1 \ll k'_c \ll \alpha^{-2},$$
which is easily accommodated by our estimates (e.g. we can have $k_c
=O(\alpha^{-1/3})).$ Thus we can assume for simplicity that $\chi$
is fixed.

\textbf{Resonances}. We define the resonances for the Hamiltonian
$H^{SM}_g$
as follows. Consider the dilations of particle positions and of
photon momenta:
$$x_j\rightarrow e^\theta x_j\  \mbox{and}\ k\rightarrow e^{-\theta} k,$$
where $\theta$ is a real parameter. Such dilations are represented
by the one-parameter group of unitary operators, $U_{\theta},$ on
the total Hilbert space ${\mathcal H}:= {\mathcal H}_p \otimes
{\mathcal F}$ of the system (see Section \ref{sec-III}). Now, for
$\theta \in \mathbb{R}$ we define the deformation family
\begin{equation}
H_{g \theta}^{SM} := X_{\theta} H_{g}^{SM} X_{\theta}^{-1} \ ,
\label{I.5}
\end{equation}
where $X_{\theta} := U_{\theta}e^{-ig F} $ with $F$, the
self-adjoint operator defined in Section \ref{sec-II}. The
transformation $H_{g}^{SM} \rightarrow e^{-ig F} H_{g}^{SM} e^{ig
F}$ is a generalization of the well-known Pauli-Fierz
transformation. Note that the operator-family $X_{\theta}$ has the
following two properties needed in order to establish the desired
properties of the resonances:

(a) $X_{\theta}$ are unitary for $\theta \in \mathbb{R}$ and

(b) $X_{\theta_1 +\theta_2}= U_{\theta_1}X_{\theta_2}$ where
$U_{\theta}$ are unitary for $\theta \in \mathbb{R}$.

It is easy to show (see Section \ref{sec-III}) that, due to
Condition (DA), the family $H_{g \theta}^{SM}$  has an analytic
continuation in $\theta$ to the disc $D(0, \theta_0)$, as a type A
family in the sense of Kato (\cite{Kato}). A standard argument shows
that the real eigenvalues of $H_{g \theta}^{SM},\ \rIm \theta
>0,$ coincide with eigenvalues of $H_{g}^{SM}$ and that complex
eigenvalues of $H_{g \theta}^{SM},\ \rIm \theta >0,$ lie in the
complex half-plane $\mathbb{C}^-$. We show below that the complex
eigenvalues of $H_{g \theta}^{SM},\ \rIm \theta
>0,$ are locally independent of $\theta$. We call such eigenvalues
the \textit{resonances} of $H_{g }^{SM}$.


As it is clear from the definition, the notion of resonance extends
that of eigenvalue and under small perturbations embedded
eigenvalues turn generally into resonances. Correspondingly, the
resonances share two 'physical' manifestations of eigenvalues, as
poles of the resolvent and and frequencies of time-periodic and
spatially localized solutions of the time-dependent Schr\"odinger
equation, but with a caveat. To explain the first property, we use
the Combes argument which goes as follows.
By the unitarity of
$X_{\theta}:= U_{\theta}e^{-igF}$ for real $\theta$,
\begin{equation}
\langle\Psi, \  (H^{SM}_{g } -z)^{-1}\Phi\rangle  = \langle
\Psi_{\bar\theta} , ( H^{SM}_{g \theta} -z)^{-1}\Phi_{\theta}
\rangle \ , \label{I.7}
\end{equation}
where $\Psi_{\theta}=X_{\theta}\Psi$, etc., for $\theta\in
\mathbb{R}$ and $z\in \mathbb{C}^+$. Assume now that $\Psi_{\theta}$
and $\Phi_{\theta}$ have analytic continuations into a complex
neighbourhood of $\theta=0$. Then the r.h.s. of \eqref{I.7} has an
analytic continuation in $\theta$ into a complex neighbourhood of
$\theta=0$. Since \eqref{I.7} holds for real $\theta$, it also holds
in the above neighbourhood. Fix $\theta$ on the r.h.s. of
\eqref{I.7}, with ${\rm Im}\theta>0$. The r.h.s.~of \eqref{I.7} can
be analytically extended across the real axis into the part of the
resolvent set of $H^{SM}_{g \theta}$ which lies in
$\overline{\mathbb{C}^-}$ and which is connected to $\mathbb{C}^+$.
This yields an analytic continuation of the l.h.s. of \eqref{I.7}.
The real eigenvalues of $H^{SM}_{g \theta}$ give real poles of the
r.h.s. of \eqref{I.7} and therefore they are the eigenvalues of
$H^{SM}_{g}$. The complex eigenvalues of $H^{SM}_{g \theta}$, which
are at the resonances of $H^{SM}_{g}$, yield the complex poles of
the r.h.s. of \eqref{I.7} and therefore they are poles of the
meromorphic continuation of the l.h.s. of \eqref{I.7} across the
spectrum of $H^{SM}_{g}$ onto the second Riemann sheet. This pole
structure is observed physically as  bumps in the scattering
cross-section or poles in the scattering matrix. There are some
subtleties involved which we explain below.


The second manifestation of resonances  alluded to above is as
metastable states (metastable attractors of system's dynamics).
Namely, one expects that the ground state is asymptotically stable
and the resonance states are (asymptotically) metastable, i.e.
attractive for very long time intervals.
%
%
More specifically, let $z_*,\ \rIm z_* \le0,$ be the ground state or
resonance eigenvalue. One expects that for an initial condition,
$\psi_0$, localized in a small energy interval around ground state
or resonance energy, $\rRe z_*$,  the solution, $\psi$, of the
time-dependent Schr\"odinger equation, $i\partial_t\psi=
H^{SM}_g\psi$,  is of the form
\begin{equation} \label{ResonDecay}
e^{-i H^{SM}_g t}\psi_0 = e^{-i z_* t}\phi_* +
O_{\textrm{loc}}(t^{-\alpha})+ O_{\textrm{res}}(g^{\beta}),
\end{equation}
for some $\alpha,\ \beta >  0$ (depending on $\psi_0$).
%
Here $\phi_*$ is either the ground state (if $z_*$ is the ground
state energy) or an excited state of the unperturbed system (if
$z_*$ is a resonance eigenvalue), the error term
$O_{\textrm{loc}}(t^{-\alpha})$ satisfies $\|(\one+|T|)^{-\nu}
O_{\textrm{loc}}(t^{-\alpha})\| \le C t^{-\alpha}$, where $T$ is the
generator of the group $U_\theta$, with an appropriate $\nu>0$, and
the error term $O_{\textrm{res}}(g^{\beta})$ is absent in the ground
state case. The reason for the latter is that, unlike bound states,
there is no 'canonical' notion of the resonance state.

The asymptotic stability of the ground state is equivalent to the
statement of local decay. Its proof was completed recently in \cite{
FroehlichGriesemerSigal2008a,  FroehlichGriesemerSigal2008c} (see
\cite{ BachFroehlichSigal1999,BachFroehlichSigalSoffer1999} for
complementary results). A statement involving survival probabilities
of excited states which is related to the metastability of the
resonances is proven in \cite{AFFS} using the results of this paper
(see \cite{HaslerHerbstHuber2007} for related results and
\cite{BachFroehlichSigal1999, Mueck, King} for partial results).

The dynamical picture of the resonance described above implies that
the imaginary part of the resonance eigenvalue, called the resonance
width, can be interpreted as the decay rate probability, and its
reciprocal, as the life-time, of the resonance.

\textbf{Main results}. Let $\epsilon^{(p)}_i$'s be the isolated
eigenvalues of the particle Hamiltonian $H_p$. 
In what follows we fix an energy $ \nu \in (\epsilon^{(p)}_0 , \inf
\sigma_{ess} (H_p))$ below the ionization threshold $\inf
\sigma_{ess} (H_p)$ and denote $\epsilon^{(p)}_{gap}\equiv
\epsilon^{(p)}_{gap}(\nu):=\min \{|\epsilon^{(p)}_i
-\epsilon^{(p)}_j |\ |\ i\neq j,\ \epsilon^{(p)}_i, \epsilon^{(p)}_j
\le \nu \}$ and $j(\nu):= \max\{j: \epsilon^{(p)}_{j} \le \nu\}$.
%

We now state the main results of this paper. 
%
\begin{theorem} \label{thm-main} Assume Condition (V). Fix $e^{(p)}_0 < \nu < \inf \sigma_{ess} (H_p)$ and let
$g\ll \min(\epsilon^{(p)}_{gap}(\nu),
\sqrt{\epsilon^{(p)}_{gap}(\nu)\tan (\theta_0/2)}\ )$. Then


(i) Each eigenvalue, $\epsilon^{(p)}_{j}$, of $H^{SM}_{g=0}$, which
is less than $\nu$,  turns into resonance and/or bound state
eigenvalues, $\epsilon_{j, k}$, of $H^{SM}_g,\ g \ne 0$;

(ii) $\forall j$, $\epsilon_{j,k} = \epsilon^{(p)}_{j} +O(g^2)$ and the total multiplicity of $\epsilon_{j, k}\
\forall k$ equals  the multiplicity of
$\epsilon^{(p)}_{j}$; 

(iii) $H^{SM}_g$ has a ground state, originating from a ground state
of $H^{SM}_{g=0}$;



(iv) $\epsilon_{j}$'s are independent of $\theta$, provided $\rIm
\theta \ge \theta_0/2$.



The statements concerning the excited states are proven under
additional Condition (DA).
\end{theorem}

%
%

In what follows we omit the subindex $k$ in $\epsilon_{j,k}$ and
write $\epsilon_{j}$. By statement (ii), we have $\epsilon_{0}:=\inf
\sigma (H^{SM}_g)$.
Let \begin{equation} \label{Sj}
S_j:=\{ z \in e^{-\theta}Q_j \mid
\rRe (e^{\theta} (z - \epsilon_j))\ge 0\
 \hbox{and}\ | \rIm (e^{\theta} (z - \epsilon_j)) |
\le \frac{1}{2} | \rRe (e^{\theta} (z - \epsilon_j)) | \}.
\end{equation}
Information about meromorphic continuation of the matrix elements of
the resolvent and position of the resonances is given in the next
theorem.
\begin{theorem} \label{thm-main2}
Assume $g\ll \epsilon^{(p)}_{gap}(\nu)$ and Conditions (V) and (DA).
Then for a dense set (defined in \eqref{D} below)
of vectors $\Psi$ and $\Phi$, the matrix elements $F(z, \Psi,
\Phi):=\langle \Psi, (H^{SM}_g-z)^{-1}\Phi\rangle$ of the resolvent
of $H^{SM}_g$ have meromorphic continuations from $\mathbb{C}^+$
across the interval $(\epsilon_{0}, \nu)$ of the essential spectrum
of $H^{SM}_g$ into the domain $\{z \in \mathbb{C}^- |\ \epsilon_{0}
< \rRe z < \nu$\},
with the wedges $S_j,\ j \le j(\nu)$, deleted. Furthermore, this
continuation has poles at $\epsilon_{j}$ in the sense that $\lim_{z
\rightarrow \epsilon_j}(\epsilon_j -z) F(z, \Psi, \Phi)$ is finite
and, for a finite-dimensional subspace of  $\Psi$'s and $\Phi$'s,
nonzero.
\end{theorem}
\textit{Discussion}.

(i) Condition (DA) could be weakened considerably so that it is
satisfied by the potential of a molecule with fixed nuclei (cf.
\cite{HunzikerSigal}).
%
%

(ii) Generically, excited states  turn into the resonances, not
bound states. A condition which guarantees that this happens is the
Fermi Golden Rule (FGR) (see \cite{BachFroehlichSigal1999}). It
expresses the fact that the coupling of unperturbed embedded
eigenvalues of  $H^{SM}_{0}$ to the continuous spectrum is effective
in the second order of the perturbation theory. It is generically
satisfied.

(iii) With little more work one can establish an explicit
restriction on the coupling constant $g$ in terms of the particle
energy difference $e^{(p)}_{gap}$ and appropriate norms of the
coupling functions.

(iv) The second theorem implies the absolute continuity of the
spectrum and its proof gives also the limiting absorption principle
in  the interval $(\epsilon_{0}, \nu)$, but these results have
already been proven by the spectral deformation and commutator
techniques \cite{
BachFroehlichSigal1999,BachFroehlichSigalSoffer1999,
FroehlichGriesemerSigal2008a}.

(v) The meromorphic continuation in question is constructed in terms
of matrix elements of the resolvent of a complex deformation,
$H^{SM}_{g, \theta},\ \rIm \theta > 0,$ of the Hamiltonian
$H^{SM}_g$.

(vi) The proof of Theorem \ref{thm-main} gives  fast convergent
expressions in the coupling constant $g$ for the ground state energy
and resonances.


The main new result of this work is the existence of resonances and
an algorithm for their computation.


The dense set mentioned in the Theorem \ref{thm-main2} is defined as
\begin{equation} \label{D}
\cD :=\bigcup_{n >0,a >0} \Ran \big(\chi_{N \le n}\chi_{|T| \le
a}\big). \end{equation} Here $N  =  \int d^3 k  a^*(k) a(k)$ is the
photon number operator and, recall, $T$ denotes the self-adjoint
generator of the one-parameter group $U_{\theta},\ \theta \in
\mathbb{R}$ (see Section \ref{sec-III}). Since $N$ and  $T$ commute,
this set is dense. We claim that for any $\psi \in \cD$, the family
$U_{\theta}e^{-ig F(x)}\psi $ has an analytic continuation from
$\mathbb{R}$ to the complex disc $D(0,\theta_0)$. Indeed, by the
construction in the next section, the family $F_{\theta}(x) : =
U_{\theta} F(x) U_{\theta}^{-1}$  has an analytic continuation from
$\mathbb{R}$ to the complex disc $D(0,\theta_0)$.
For $\theta$ complex this continuation is a family of
non-self-adjoint operators. However, the exponential $e^{-ig
F_{\theta}(x)}$ is well defined on the dense domain
$\bigcup_{n<\infty} \Ran \chi_{N \le n}$.
Since $$U_{\theta} e^{igF(x)}\psi= e^{igF_{\theta}(x)}\chi_{N \le
n}U_{\theta}\chi_{|T| \le a}\psi$$ for some $n$ and $a$, s.t.
$\chi_{N \le n}\chi_{|T| \le a}\psi= \psi$, the family
$U_{\theta}e^{-ig F(x)}\psi$ has an analytic continuation in
$\theta$ from $\mathbb{R}$ to $D(0,\theta_0)$.
%
%

\textbf{Infrared problem}.
%
%
As is shown in Theorem \ref{thm-main} and is understood in Phyisics
on the basis of formal - but rather non-trivial - perturbation
theory, the resonances arise from the eigenvalues of the free
Hamiltonian $H^{SM}_0$. To find the spectrum of $H^{SM}_0$ one
verifies that $\hf$ defines a positive, self-adjoint operator on
$\cF$ with purely absolutely continuous spectrum, except for a
simple eigenvalue $0$ corresponding to the vacuum eigenvector $\Om$
(see Supplement A). Thus, for $g=0$ the low energy spectrum of the
operator $H^{SM}_0$ consists of branches $[\epsilon^{(p)}_i,
\infty)$ of absolutely continuous spectrum and of the eigenvalues
$\epsilon^{(p)}_i$'s, sitting at the continuous spectrum
'thresholds' $\epsilon^{(p)}_i$'s.
The absence of gaps between the eigenvalues and thresholds is a
consequence of the fact that the photons  are massless. This leads
to hard problems in perturbation theory, known collectively as the
\textit{infrared problem}.

This situation is quite different from the one in Quantum Mechanics
(e.g. Stark effect or tunneling decay) where the resonances are
isolated eigenvalues of complexly deformed Hamiltonians. This makes
the proof of their existence and establishing their properties, e.g.
independence of $\theta$ (and, in fact, of the transformation group
$X_{\theta}$), relatively easy. In the non-relativistic QED (and
other massless theories),
giving meaning of the resonance poles and proving independence of
their location of $\theta$ is a rather involved matter (see below).

The point above can be illustrated on the proof of the statement
\eqref{ResonDecay}. To this end we use
the formula
\begin{align*}
& e^{-i H t}f(H)= \frac{1}{\pi} \int_{-\infty}^\infty d\lambda
f(\lambda)e^{-i\lambda t}\rIm(H-\lambda-i0)^{-1}
\end{align*}
(see e.g. \cite{RSIV}) connecting the propagator and the resolvent.
For the ground state the absolute continuity of the spectrum outside
the ground state energy, or a stronger property of the limiting
absorption principle, suffices to establish the result in question.
In the resonance case, one uses  the fact that the meromorphic
continuation of matrix elements of the resolvent (on an appropriate
dense set of vectors) to the second Riemann sheet has poles at
resonances and performing a suitable deformation of the contour of
integration in the formula above (see e.g. \cite{Hunziker}). This
works
when the resonances are isolated (see \cite{Hunziker,
HunzikerSigal}). In the present case,
proving \eqref{ResonDecay} is a subtle problem.

\textbf{Resonance poles}.
Can we make sense of the resonance poles in the present context? The
answer to this question is obtained in \cite{AFFS}, where it is
shown, assuming the results of this paper, that
for each $\Psi$ and $\Phi$ from a dense set of
vectors, the meromorphic continuation, $F(z, \Psi, \Phi)$, of the
matrix element $\langle \Psi, (H^{SM}_g-z)^{-1}\Phi\rangle$,
described above, is of the following form:
%
%
%
%
%
%
%
%
%
\begin{equation} \label{poles}
F(z, \Psi, \Phi)=(\epsilon_j -z)^{-1} p(\Psi, \Phi) + r(z, \Psi,
\Phi) \comma
\end{equation}
%
 near the resonance
$\epsilon_j$ of $H^{SM}_{g }$. Here $p$ and $r(z)$ are sesquilinear
forms in $\Psi$ and $\Phi$ with $r(z)$, analytic in $z \in Q:= \{z
\in \mathbb{C}^- |\ \epsilon_{0} < \rRe z < \nu \} / \bigcup_{j \le
j(\nu)} S_j $ and bounded on the intersection of a neighbourhood of
$\epsilon_j$ with $Q$ as $$|r(z, \Psi, \Phi)| \le C_{\Psi,
\Phi}|\epsilon_j-z|^{-\gamma}\ \mbox{for some}\ \gamma <1.$$
Moreover, $p \ne 0$ at least for one pair of vectors $\Psi$ and
$\Phi$ and $p = 0$
for a dense set of vectors $\Psi$ and $\Phi$ in a finite
co-dimension subspace.
The multiplicity of a resonance is the rank of the residue at the
pole.
%
\textit{The next important problem is to connect the ground state
and resonance eigenvalues to poles of the scattering matrix}.

%
%


\textbf{Approach}.
To prove Theorems \ref{thm-main} and \ref{thm-main2} we apply the
spectral renormalization group (RG) method
(\cite{BachChenFroehlichSigal2003,
BachFroehlichSigal1998a,BachFroehlichSigal1998b, GriesemerHasler2,
FroehlichGriesemerSigal2008b}) to the Hamiltonians $ e^{-ig F}
H_{g}^{SM} e^{ig F}$ (the ground state case) and $H_{g \theta}^{SM}$
(the resonance case).
Note that the version of RG needed in this work uses new --
anisotropic -- Banach spaces of operators, on which the
renormalization group acts. It is described in \cite{
FroehlichGriesemerSigal2008b}.
Using the RG technique we describe the spectrum of the operator
$H^{SM} _{g \theta}$ in $\{z \in \mathbb{C}^- |\ \epsilon_{0} < \rRe
z < \nu$\} from which we derive Theorems \ref{thm-main} and
\ref{thm-main2}.

%
%

In the terminology of the Renormalization Group approach the
perturbation in \eqref{Hsm} is marginal (similar to critical
nonlinearities in nonlinear PDEs). This leads to the presence of the
second zero eigenvalue in the spectrum of the linearized RG flow
(note that there is no spectral gap in the linearized RG flow). This
case is notoriously hard to treat as one has to understand the
dynamics on the implicitly defined central manifold. The previous
works \cite{BachFroehlichSigal1998a,BachFroehlichSigal1998b} remove
it by either assuming the non-physical infrared behaviour of the
vector potential by replacing $|k|^{-1/2}$ in the vector potential
\eqref{A} by $|k|^{-1/2 +\varepsilon}$, with $\varepsilon
>0$ or or by assuming presence of a strong confining external potential
so that $V(x) \ge c|x|^2$ for $x$ large. Our work shows that in
non-relativistic QED one can overcome this problem by suitable
canonical transformation and choice of the Banach space.


%


Our approach is also applicable to Nelson's model describing
interaction of particles with massless lattice excitations (phonons)
described by a quantized, massless, Boson field (see Supplement B)
and Theorems \ref{thm-main} and \ref{thm-main2} are  still valid if
replace there the operator $H^{SM}_g$ by  the Hamiltonian $H^{N}_g$
for this model. (In this case we recover earlier results.)
%
%
In fact, we consider a class of generalized particle-field operators
(introduced in Section \ref{sec-IV}) which contains both, operators
$H_g^{PF}$ and $H_g^{N}$.

\textbf{Organization of the paper}. The paper is organized as
follows. In Section \ref{sec-II} we introduce the generalized
Pauli-Fierz transformation ($H_{g}^{SM}  \rightarrow e^{-ig F}
H_{g}^{SM} e^{ig F}=: H_{g}^{PF}$ ) and in Section \ref{sec-III},
the complex deformation of quantum Hamiltonians. In Section
\ref{sec-IV} we introduce a class of generalized particle-field
Hamiltonians and show that the Hamiltonian $H_{g}^{PF}$ obtained in
Section \ref{sec-II} and the Hamiltonian $H_{g}^{N}$  as well as
their dilation deformations belong to this class. In the rest of the
paper we study the Hamiltonians from the class introduced and derive
Theorems \ref{thm-main} and \ref{thm-main2} from the results about
these Hamiltonians. In Section \ref{sec-V} we introduce an
isospectral Feshbach-Schur map and use it to map the generalized
particle-field Hamiltonians into Hamiltonians acting only on the
field Hilbert space - Fock space (elimination of particle and high
photon energy degrees of freedom). The image of this map is shown in
Section \ref{sec-VII} to belong to a certain neighbourhood in the
Banach spaces introduced in Section \ref{sec-VI}. The latter spaces
are an anisotropic - in the momentum representation - modification
of the Banach spaces used in \cite{BachChenFroehlichSigal2003,
BachFroehlichSigal1998a,BachFroehlichSigal1998b}. In Section
\ref{sec-VIII} we use the results of
\cite{FroehlichGriesemerSigal2008b} on the spectral renormalization
group (cf. \cite{BachChenFroehlichSigal2003,
BachFroehlichSigal1998a,BachFroehlichSigal1998b}) to describe the
spectrum of generalized particle-field Hamiltonians.
Finally, in Section \ref{sec-IX} we prove Theorems \ref{thm-main}
and \ref{thm-main2}. In Appendix A we recall some properties of the
Feshbach-Schur map and in Appendix B we prove the main result of
Section \ref{sec-VI}. The results of both appendices are close to
certain results from  \cite{BachChenFroehlichSigal2003, GriesemerHasler2,
FroehlichGriesemerSigal2008b}, but there
are a few important differences. The main ones are that we have to
deal with unbounded interactions and, more importantly, with
momentum-anisotropic spaces. Some basic facts about Fock spaces and
creation and annihilation operators on them are collected in
Supplement A and in Supplement B we describe the Nelson Hamiltonians
and their dilation deformations.


\secct{Generalized Pauli-Fierz transformation} \label{sec-II}
%
In order to simplify notation from now on we assume that the number
of particles is $1$, $n=1$. We also set the particle mass to $1$,
$m=1$. The generalizations to an arbitrary number of particles is
straightforward. We define the generalized Pauli-Fierz
transformation mentioned in the introduction: with $F(x)$ introduced
below we let
\begin{equation} \label{HPF}
H_g^{PF}: = e^{-ig F(x)} H^{SM}_g e^{ig F(x)}.
\end{equation}
We call the resulting Hamiltonian
the generalized Pauli-Fierz Hamiltonian. Here $F(x)$ is the
self-adjoint operator on the state space $\cH$ given by
%
\begin{equation}\label{eq3}
F(x)=\sum_\lambda
\int(\bar{f}_{x,\lambda}(k)a_{\lambda}(k)+f_{x,\lambda}(k)a_{\lambda}^*(k))\frac{d^3k}{\sqrt{|k|}},
\end{equation}
with the coupling function $f_{x,\lambda}(k)$ chosen as
\begin{equation}\label{f}f_{x,\lambda}(k):=
e^{-ikx}\frac{\chi(k)}{\sqrt{|k|}}\varphi(|k|^{\frac{1}{2}}e_\lambda(k)
\cdot x).
\end{equation} The function $\varphi$ is assumed to be $C^2$, bounded, with bounded second derivative, and
satisfying $\varphi'(0)=1.$ We assume also that  $\varphi$ has a
bounded analytic continuation into the wedge $\{z \in \mathbb{C} |\
|\arg(z)| < \theta_0\}$. We compute


\begin{equation} \label{H^PFa}
H_g^{PF} = \frac{1}{2} (p - g A_1(x))^2 + V_g(x) + H_f + gG(x)\\
\end{equation}
where $A_1(x) = A(x) - \nabla F(x),\ V_g(x):= V(x) + 2
g^2\sum_\lambda \int |k| |f_{x,\lambda}(k)|^2d^3k$ and
\begin{equation} \label{I.14}
G(x):=- i\sum_\lambda
\int|k|(\bar{f}_{x,\lambda}(k)a_{\lambda}(k)-f_{x,\lambda}(k)a_{\lambda}^*(k))
\frac{d^3k}{\sqrt{|k|}}.
\end{equation}
%
%
(The terms $gG$ and $V_g - V$ come from the commutator expansion
$e^{-ig F(x)} H_f e^{ig F(x)}$ $= - i g [F,H_f] - g^2 [F, [F,
H_f]]$.) Observe that the operator-family $A_1(x)$ is of the form
\begin{equation}\label{15}
A_1(x)=\sum_\lambda
\int(e^{ikx}a_{\lambda}(k)+e^{-ikx}a_{\lambda}^*(k))\chi_{\lambda,
x}(k){d^3k\over \sqrt{|k|}},
\end{equation}
where the coupling function, $$\chi_{\lambda, x}(k):=
e_{\lambda}(k)e^{-ikx}\chi(k)- \nabla_x f_{x,\lambda}(k),$$
satisfies the estimates
\begin{equation}\label{chi-estim1}
|\chi_{\lambda, x}(k)|
\le \const \min (1, \sqrt{|k|}\la x\ra),
\end{equation}
with $\la x\ra := (1 +|x|^2)^{1/2}$, and
\begin{equation}\label{chi-estim2}
\int \frac{d^3 k}{|k| } \: |\chi_{\lambda, x}(k)|^{2}  \ < \ \infty
.
\end{equation}
The fact that the operators $A_1$ and $G$ have better infra-red
behavior than the original vector potential $A$, is used in proving,
with a help of a renormalization group, the existence of the ground
state and resonances for the Hamiltonian $H_g^{SM}$.

We note that for the standard  Pauli-Fierz transformation the
function $f_{x,\lambda}(k)$ is chosen to be $\chi(k)e_\lambda(k)
\cdot x$, which results in the operator $G$ (which in this case is
proportional to the electric field at $x=0$ dot $x$) growing as $x$.

We mention for further references that the operator (I.13) can be
written as
\begin{equation} \label{Hpf}
H_g^{PF} \ = \ H_{0}^{PF} \, + \, I_{g}^{PF} \comma
\end{equation}
where $H^{PF}_{0}=H_0 + 2 g^2\sum_\lambda \int
|k||f_{x,\lambda}(k)|^2d^3k + g^2\sum_\lambda
\int{|\chi_\lambda(k)|^2\over |k|}d^3k$, with $H_0 := H_p+H_f$
and $I_{g}^{PF}$ is defined by this relation. Note that the operator
$I_{g}^{PF}$ contains linear and quadratic terms in the creation and
annihilation operators and that the operator $H^{PF}_{0}$ is of the
form $H^{PF}_{0}= H^{PF}_{p}+H_{f}$ where
\begin{equation} \label{Hpg}
H^{PF}_{p}:=H_{p}+2 g^2\sum_\lambda \int |k|
|f_{x,\lambda}(k)|^2d^3k + g^2\sum_\lambda
\int{|\chi_\lambda(k)|^2\over |k|}d^3k
\end{equation}
with $H_p$ given in \eqref{Hp}.

%
Since the operator $F(x)$ in \eqref{HPF} is self-adjoint, the
operators $H_g^{SM}$ and $H_g^{PF}$ have the same eigenvalues with
closely related eigenfunctions and the same essential spectra.
%




\secct{Complex Deformation and Resonances} \label{sec-III}
%

In this section we define complex transformation of the Hamiltonian
under consideration which underpins the proof of the resonance part
of Theorem \ref{thm-main} and the proof of Theorem \ref{thm-main2}.
%
Let $u_{ \theta}$ be the dilatation transformation on the one-photon
space, i.e., $u_{ \theta}$: $f(k)\to e^{-3\theta/2}
f(e^{-\theta}k)$.
Define the dilatation transformation, $U_{f \theta}$, on the Fock
space, $\cH_{f}\equiv \cF$, by second quantizing $u_{ \theta}$:
$U_{f \theta}: = e^{iT \theta}$ where $T: = \int a^*(k) t a(k) dk$
and $t$ is the generator of the group $u_{ \theta}$ (see Supplement
for the careful definition of the above integral). This gives, in
particular,
\begin{equation}
U_{f \theta} a^* (f)  U_{f \theta}^{-1}  = a^* (u_{ \theta} f) \ .
\label{III.1}
\end{equation}
Denote by $U_{p \theta}$ the standard dilation group on the particle
space: $U_{p \theta} : \psi (x) \to e^{{3\over 2} \theta} \psi
(e^\theta x)$ (remember that we assumed that the number of particles
is $1$). We define the dilation transformation on the total space
$\cH=\cH_{p}\otimes\cH_{f}$ by
\begin{equation}
U_{ \theta}  =  U_{p \theta}  \otimes U_{f \theta}. \label{III.2}
\end{equation}

For $\theta \in \mathbb{R}$ the above operators are unitary and we
can define the family of Hamiltonians originating from the
Hamiltonian $H_{g}^{SM}$ as
\begin{equation}
H_{g \theta}^{SM} := U_{ \theta}e^{-igF(x)} H_{g}^{SM} e^{igF(x)}U_{
\theta}^{-1} \ . \label{III.3}
\end{equation}
Under Condition (DA), there is a Type-A (\cite{Kato}) family
$H^{SM}_{g \theta}$ of operators analytic in the domain $ |\rIm
\theta| < \theta_0$, which is equal to \eqref{III.3} for $\theta \in
\mathbb{R}$ and s.t. $H^{SM
*}_{g \theta} = H^{SM}_{g \overline{\theta}}$ and
\begin{equation}
H^{SM}_{g \theta}  =  U_{\rRe\theta}  H^{SM}_{g  i\rIm\theta}
U_{\rRe\theta}^{-1}. \label{III.4}
\end{equation}
Indeed, using the decomposition $H^{PF}_g = H^{PF}_p+H_f
+I_{g}^{PF}$ (see \eqref{Hpf}-\eqref{Hpg}),
we write for $\theta \in \mathbb{R}$
\begin{equation}
H_{g \theta}^{SM} = H_{p\theta}^{SM}    \otimes \mathbf{1}_f +
e^{-\theta} \mathbf{1}_{p} \otimes H_f + I_{g \theta}^{SM} \ ,
\label{III.5}
\end{equation}
where $H_{p \theta}^{SM} := U_{p \theta} H_{p}^{PF} U_{p
\theta}^{-1}$ and $I_{g \theta}^{SM} := U_{ \theta} I_{g}^{PF} U_{
\theta}^{-1}$.
It is not hard to compute that $H_{p \theta}^{SM} = -e^{-2\theta}
{1\over 2} \Delta + V_g(e^\theta x)$, where
\begin{equation} \label{Vg}
V_g( x):=V(x)+2 g^2\sum_\lambda \int |k| |f_{x,\lambda}(k)|^2d^3k +
g^2\sum_\lambda \int{|\chi_\lambda(k)|^2\over |k|}d^3k
\end{equation}
with $V$ given in \eqref{Hp}. Furthermore, using \eqref{III.1} and
the definitions of the interaction $I_{g}^{PF}$, we see that $I_{g
\theta}^{SM} $ is obtained from $I_{g}^{PF}$ by the replacement
$a^\#(k)\ \rightarrow e^{-{3\theta\over 2}}a^\#(k)$ and, in the
coupling functions only,
\begin{equation} k\ \rightarrow e^{-\theta}k\ \mbox{and}\  x\ \rightarrow
e^{\theta}x.
\label{III.7}
\end{equation}
By Condition (DA), the family \eqref{III.5} is well defined for all
$\theta$ satisfying $ |\rIm \theta| < \theta_0$ and has all the
properties mentioned after Eqn \eqref{III.3}. Hence, for these
$\theta$, it gives the required analytic continuation of
\eqref{III.3}.
We call $H^{SM}_{g \theta}$ with $\rIm \theta
>0$ the complex deformation of $H^{SM}_{g}$.

Recall that we define the the resonances of $H^{SM}_g$ as the
complex eigenvalues of $H^{SM}_{g \theta}$ with $\rIm \theta
>0$.
Thus to find resonances (and eigenvalues)
of $H^{SM}_{g }$
we have to locate complex (and real) eigenvalues of $H^{SM}
_{g\theta}$
for some $\theta$ with $\rIm \theta > 0$.

In Sections \ref{sec-V} - \ref{sec-VIII} we prove the following
result
%
\begin{theorem} \label{thm-specHtheta} Assume Conditions (V) and (DA) holds. Fix $e^{(p)}_0 < \nu < \inf \sigma_{ess}
(H_p)$ and let $g\ll \epsilon^{(p)}_{gap}(\nu)$. Then the operators
$H^{SM}_{g \theta}$, with $\rIm \theta >0$, have eigenvalues
$\epsilon_{j},\ j \le j(\nu),$ s.t.  $\epsilon_{j} =
\epsilon^{(p)}_{j} +O(g^2)$ and $\epsilon_{j}$ are independent of
$\theta$. The essential spectrum of $H^{SM}_{g \theta},\ \rIm \theta
>0,$ is a subset of the set $\bigcup_{j \le j(\nu)} S_j,$ where the
sets $S_j$ are given in \eqref{Sj}.
\end{theorem}
Theorem  \ref{thm-specHtheta}, together with the discussion in
paragraphs containing Eqns \eqref{D}- \eqref{I.7} implies Theorems
\ref{thm-main} and \ref{thm-main2} (for the ground state part of
Theorem \ref{thm-main} it contains unnecessary Condition (DA)).

Furthermore, one can show that the eigenvalues $\epsilon_{j},\ j \le
j(\nu),$ have the properties


(i) If the FGR condition is satisfied, then $\rIm \epsilon_{j} =
-g^2\gamma_{j} +O(g^4)$, where $\gamma_{j}$ are given by the Fermi
Golden Rule formula;

(ii) $\epsilon_{j}$  can be computed explicitly in terms of fast
convergent expressions in the coupling constat $g$.
\secct{Generalized Particle-Field Hamiltonians} \label{sec-IV}
%
It is convenient to consider a more general class of Hamiltonians
which contains, in particular, both, the generalized Pauli-Fierz and
Nelson Hamiltonians and their complex dilation transformations.
(Recall that the Nelson hamiltonian is defined in Supplement B.) We
consider Hamiltonians of the form
\begin{equation}\label{Hg}
H_g \ = \ H_{0g} \, + \, I_g,
\end{equation}
where $g>0$ is coupling constant, $H_{0g}:=H_{pg} + H_f$, with
$H_{pg}:=- \kappa \Delta+V_g(x),\ \kappa \in \mathbb{C},\ \kappa \ne
0,$ and $I_g:= g\sum_{1 \le m+n \leq 2} W_{m,n}$. We assume that
$V_g(x)$ is $\Delta-$bounded with the relative bound less than
$|\kappa|/2$, more precisely, that it obeys the bound
\begin{equation}\label{Vbnd}
\| V_g\psi \| \le \frac{|\kappa|}{2}\|\Delta \psi\| + \|\psi\|,
\end{equation}
uniformly in $g \le1$, where we set the constant in front of the
second term on the r.h.s. to $1$. This constant plays no role in our
analysis. Moreover, we assume that the operators $W_{m,n}$ are of
the form
\begin{eqnarray}\nonumber
W_{m,n}&&:= \int_{\mathbb{R}^{3(m+n)}} \prod_{1}^{m+n}(\frac{ dk_{j}
}{ |k_{j}|^{1/2} }) \; \prod_{1}^{m}a^*( k_{j} ) \,
\\ \label{Imn}
&& \times w_{m,n} \big[k_{1}, ...,k_{m+n}  \big] \,
\prod_{m+1}^{m+n}a( k_{j} ) \: ,
\end{eqnarray}
where $w_{m,n}[k],\ k:=(k_{1}, ...,k_{m+n})$, the coupling
functions, are operator-functions from $\mathbb{R}^{3(m+n)}$ to
bounded operators on the particle space $\cH_{p}$ obeying
\begin{equation}\label{Ibnd}
\sup_{g \le 1} \|w_{m,n}\|^{(0)}_{\mu} <\infty,
\end{equation}
for some $\mu \ge 0$ and $\delta_0 >0$ (the latter parameter is not
displayed, see the next equation; also note that $w_{m,n}$ might
depend on the coupling constant $g$). Here the norm
$\|w_{m,n}\|^{(0)}_{\mu}$ is defined by
%
%
%
%
%
%
%
\begin{equation}\label{Inorm}
\|w_{m,n}\|^{(0)}_{\mu}:=\sup_{|\delta| \le \delta_0}
\sup_{k \in \mathbb{R}^{3(m+n)}} \big\|\frac{e^{-\delta \langle x
\rangle}
w_{m,n}[k]e^{\delta \langle x \rangle}\langle p\rangle^{-(2-m-n)}}
{[\min ( \langle x \rangle^{m+n}\prod_{1}^{m+n}( |k_{j}|^{1/2}
),1)]^\mu}\big\|_{part}.
%
\end{equation}
Here $\|\cdot \|_{part}$ is the operator norm on the particle
Hilbert space $\cH_p$.
We observe that for $g$ sufficiently small
$$D(H_g)=D(H_{0}) \subset D(I_g).$$

We denote by $GH_\mu$ the class of (generalized particle-field)
Hamiltonians satisfying the restrictions \eqref{Hg} - \eqref{Inorm}.
We also denote by $GH^{m n}_\mu$ the class of operators of the form
\eqref{Imn} - \eqref{Inorm}.

Clearly, both, the generalized Pauli-Fierz and Nelson, Hamiltonians
belong to $GH_\mu$ with $\mu=1/2$ for the generalized Pauli-Fierz
Hamiltonian and $\mu>0$ for the Nelson Hamiltonian and with  $\kappa
= 1/2$. Indeed, for the Nelson model, ~\eqref{Hn}-~\eqref{XIII.5},
$V_g=V$ obeys \eqref{Vbnd} and $w_{m,n}$ are $0$ for $m+n=2$ and
multiplication operators by the bounded functions $\kappa(k)  e^{-i
k x}$ and $\kappa(k) e^{i k x}$ for $m+n=1$. For the QED case (the
generalized Pauli-Fierz Hamiltonian, \eqref{H^PFa}) $V_g$ is given
by \eqref{Vg}
and $I_{g}:= p \cdot A_1(x)+ \frac{1}{2}g :A_1(x)^2: + G(x)$, where
the operator $G(x)$ is defined in \eqref{I.14}. From these
expressions we see that $V_g$ satisfies \eqref{Vbnd} and $w_{m,n}$
obey the conditions formulated above.

Dilation deformed generalized Pauli-Fierz and Nelson Hamiltonian
also fit this framework.  Let $H^{SM}_{g \theta}$ be a complex
deformation of the QED Hamiltonian $H^{SM}_{g }$, i.e. the dilation
transformation of the generalized Pauli-Fierz Hamiltonian $H^{PF}_{g
}$. Then the operator $H_g : = e^{\theta} H^{SM}_{g \theta}$
satisfies the restrictions imposed above with $\mu = 1/2$ and
$\kappa = e^{-\rRe \theta}/2$. For the Nelson model we have and $\mu
>0$.
%
%
\secct{Elimination of Particle and High-Photon Energy Degrees of
Freedom} \label{sec-V}

In this section we consider the operator families $H_g-\lambda$,
where the operator $H_g=H_{g0}+I_{g}
\in GH_\mu$
(see Section \ref{sec-IV}),
%
and map them into families of operators acting on the Fock space
only (elimination of the particle degrees of freedom). We will study
properties of the latter operators in Sections \ref{sec-VII} and
\ref{sec-VIII} after we introduce appropriate Banach spaces in
Section \ref{sec-VI}.


Fix $1 \le j \le j(\mu)$ and consider an eigenvalue $\lambda_{j} \in
\sigma_d (H_{pg})$ and define
\begin{equation} \label{delta_j}
\delta_j : = \dist (\lambda_j,\sigma (H_{pg})/\{\lambda_j \}
+[0,\infty)).
%
\end{equation}
We assume $\delta_j >0$ and we
define the set
\begin{equation} \label{Q_j}
Q_j : = \{ \lambda \in \mathbb{C} \mid \rRe (\lambda - \lambda_j)
\le {1 \over 3}\delta_j\ \hbox{and}\ | \rIm (\lambda - \lambda_j)|
 \le {1 \over 3}\delta_j \}.
\end{equation}

Let $P_{pj}$ be the orthogonal projection onto the eigenspace of
$H_{pg}$ corresponding to $\lambda_j$  and, as usual, ${\overline
P_{pj}} = \mathbf{1} - P_{pj}$.  We define $H_{pg}^{\delta}
:=e^{-\varphi}H_{pg} e^{\varphi}$ and $P_{pj}^{\delta} :=
e^{-\varphi}P_{pj}\ e^{\varphi}$ with $\varphi=\delta\langle x
\rangle$. We use the following parameter to measure the size of the
resolvent of $H_{pg}^{\delta}$:
\begin{equation} \label{kappa_j}
\kappa_{j}^{-1} := \sup_{0 \le \delta \le \delta_0}\sup_{\lambda \in
Q_j} \|(H_{pg}^{\delta} - \lambda)^{-1} {\overline P_{pj}^{\delta}}
\|,
\end{equation}
for $\delta_0 >0$ sufficiently small. Note that if the operator
$H_{pg}$ is normal, as in the case of the problem of the ground
state, where $H_{pg}$ is self-adjoint, then $\kappa_j
$ can be easily estimated for $\delta_0 $ sufficiently small.
%
%
If the operator $H_{pg}$ is not normal, then getting an explicit
upper bound on its resolvent requires some work.
This will be done in the proof of Theorem \ref{thm-specHtheta} given
in Section \ref{sec-IX}.
%
%

%


Our goal now is to define the renormalization map on the class
generalized particle-field Hamiltonians $GH_\mu$. This map is a
composition of three maps which we introduce now. First of these is
the smooth Feshbach-Schur map (SFM)\footnote{In
\cite{BachFroehlichSigal1998a,BachFroehlichSigal1998b,
BachChenFroehlichSigal2003} this map is called the Feshbach map. As
was pointed out to us by F. Klopp and B. Simon, the invertibility
procedure at the heart of this map was introduced by I. Schur in
1917; it appeared implicitly in an independent work of H. Feshbach
on the theory of nuclear reactions, in 1958, see
\cite{GriesemerHasler} for further extensions and historical
remarks}, or decimation, map, $F_{ \pi},$ which is defined as
follows. We introduce a pair of almost projections
\begin{equation} \label{pi}
\pi \equiv \pi_j \equiv \pi[H_f]:=P_{pj}\otimes\chi_{H_f\le\rho}
\end{equation}
and $\overline{\pi} \equiv \overline{\pi}[H_f]$ which form a
partition of unity
$\pi^{2}+ \overline{\pi}^{2} = \mathbf{1}$. Note that $\pi$ and
$\overline{\pi}$ commute with $H_{0g}$ introduced in Section
\ref{sec-IV}. Next, for $H_{g}= H_{0g} +I_{g} \in GH_\mu$, we define
\begin{equation}
\\ \label{V.5}
H_{\bpi} \ \; :=  H_{0g}\: + \: \bpi I_{g}\bpi \period
\end{equation}
Finally, on the operators $H_{g} - \lambda$ s.t. $H_{g}= H_{0g}
+I_{g} \in GH_\mu$ and
\begin{equation}
\label{V.6}H_{\bpi} - \lambda\ \mbox{is (bounded) invertible on}\
\Ran \, \bpi,
\end{equation}
we define \textit{smooth Feshbach-Schur map}, $F_{ \pi},$ as
%
%
\begin{equation} \label{Fesh}
 F_{\pi} (H_{g} - \lambda) \ := \ H_{0g} - \lambda \, + \, \pi I_{g}\pi \, -
\, \pi  I_{g} \bpi (H_{\bpi}-\lambda)^{-1} \bpi I_{g} \pi \period
\end{equation}
Observe that the last two operators on the r.h.s. are bounded since,
for any operator $I_{g}$ as described in Section \ref{sec-IV},
$$I_{g} \pi\ \mbox{and}\ \pi I_{g}\ \mbox{extend to bounded
operators on}\ \cH.$$ Properties of the smooth Feshbach-Schur maps,
used in this paper, are described in Appendix A.
For more details see
\cite{BachChenFroehlichSigal2003,GriesemerHasler}.
%
%

Next, we introduce the \textit{scaling transformation} $S_\rho:
\cB[\cH] \to \cB[\cH]$, which acts on the particle component of
$\cH:=\cH_{p}\otimes\cH_f$ by identity and on the field one, by
%
%
%
\begin{equation} \label{Srho}
S_\rho(\one) \ := \ \one \comma \hspace{5mm} S_\rho (a^\#(k)) := \
\rho^{-d/2} \, a^\#( \rho^{-1} k) \comma
\end{equation}
where $a^\#(k)$ is either $a(k)$ or $a^*(k)$ and $k \in \RR^3$.
%

Now,  on Hamiltonians acting on $\cH:=\cH_{p}\otimes\cH_f$ which are
in the domain of the decimation map $F_{\pi}$ we define the
renormalization map $\cR_{\rho j}$ as
\begin{equation}
\cR_{\rho j}=\rho^{-1} S_{\rho}\circ F_{ \pi}, \label{RGmap}
\end{equation}
where recall $\pi\equiv \pi_j$. The parameter $ \rho $ here is the
same as the one in \eqref{pi}. It gives
a photon energy scale and it is restricted below.

%
To simplify the notation we assume that the
eigenvalue $\lambda_j$ of the operator $H_{pg}$ is simple (otherwise
we would have to deal with matrix-valued operators on $\cH_f$).  We
have
\begin{theorem}
\label{thm-V.1} Let $H_g$ be a Hamiltonian of the class $GH_\mu$
defined in Section IV
%
with $ \mu \ge 0$. We assume that $\delta_j > 0$.
%
Then for $g \ll \rho\le \kappa_{j}
$ and $\lambda\in Q_j$
\begin{equation}
H_g-\lambda\in D(\cR_{\rho j}). \label{V.10}
\end{equation}
Furthermore, $\cR_{\rho j}(H_g -\lambda) = {P}_{pj}\otimes\
H_{\lambda j} + (H_{0g}-\lambda)({\bar{P}}_{pj}\otimes\ \mathbf{1})
$ where the family of operators $H_{\lambda j} $, acting on $\cF$,
is s.t. $H_{\lambda j} - \hf$ is bounded and analytic in $\lambda\in
Q_j$.
%
\end{theorem}
%
%
%
%

A proof of Theorem \ref{thm-V.1} is similar to that of related
results of
\cite{BachFroehlichSigal1998a,BachFroehlichSigal1998b,BachFroehlichSigal1999}.
%
%
%
%
We begin with
%
%
\begin{proposition} \label{prop-V.2}
Let $g \ll \rho \le \kappa_{j}$ and $\lambda\in Q_j$. Then the
operators $H_{\overline{\pi}} -\lambda$ are invertible on $\Ran
\overline{\pi}$ and we have the estimate
\begin{equation}\label{V.11a}
\|\overline{\pi}(H_{ \overline{\pi}} -\lambda)^{-1}\overline{\pi} \|
\le 4 \rho ^{-1}.
\end{equation}
\end{proposition}
\Proof
First we show that for $\lambda\in Q_j$ the operator $H_{0g}
-\lambda$ is invertible on $\Ran \overline{\pi}$  and the following
estimate holds  for $n= 0, 1$
\begin{equation} \label{V.9}
\|\big(|p|^2+H_f+1 \big)^{n} R_0 (\lambda) \| \le C\rho ^{-1}
\end{equation}
where $R_0(\lambda):=(H_{0g} -\lambda)^{-1}\overline{\pi}$.
%
If the operator $H_{pg}$ is self-adjoint then the estimates above
are straightforward. In the non-self-adjoint case we proceed as
follows.

Write $\overline{\pi} = P_{pj} \otimes \chi_{H_f \ge \rho} +
{\overline P_{pj}} \otimes \mathbf{1}$, where, as usual, ${\overline
P_{pj}} = \mathbf{1} - P_{pj}$. Since $H_{0 g} = \lambda_j + H_f\
\hbox{on Ran} (P_{pj} \otimes \chi_{H_f \ge \rho})$, the operator
$H_{0 g} - \lambda$ is invertible $\hbox{on Ran} (P_{pj} \otimes
\chi_{H_f \ge \rho})$ for $ \lambda \in Q_j$ and
\begin{equation} \label{V.10}
\| (H_{0 g} - \lambda)^{-1} (P_{pj} \otimes \chi_{H_f \ge \rho}) \|
\le 2 \rho ^{-1}\ .
\end{equation}

Next, $\sigma (H_{0g} |_ {Ran (\overline P_{pj}\otimes \mathbf{1})})
=\sigma (H_{pg} |_ {Ran \overline P_{pj}}) + \sigma (H_{f})=\sigma
(H_{pg})/\{\lambda_j\} + \bar{\mathbb{R}} ^{+}$. Now, by the
definition of $Q_j$ we have $\inf_{s\ge 0}\dist(\lambda_j -s, Q_j)
\le \delta_j/2$. This and the definition of $\delta_j$ give
\begin{equation} \label{V.11}
\dist (\sigma (H_{0g} |_ {Ran (\overline P_{pj}\otimes
\mathbf{1})}),
Q_j) \ge \delta_j /2.
\end{equation}
Therefore, for $\lambda \in Q_j,$ the operator $ H_{0 g} - \lambda$
is invertible on $\hbox{Ran} (\overline P_{pj} \otimes \mathbf{1})$.
Since the operator $(H_{0g} - \lambda)^{-1} ({\overline P_{pj}}
\otimes \mathbf{1})$ is analytic in a neighbourhood of
$\overline{Q_j}$ we have that
$\sup_{\lambda \in Q_j} \|(H_{0g} - \lambda)^{-1} ({\overline
P_{pj}} \otimes \mathbf{1}) \| < \infty.$

We claim that
\begin{equation} \label{V.12}
\sup_{\lambda \in Q_j} \|(H_{0g} - \lambda)^{-1} ({\overline P_{pj}}
\otimes \mathbf{1}) \| \le C \kappa_{j}^{-1}
\end{equation}
where $\kappa_{j}$ is defined in \eqref{kappa_j}. Indeed, since the
operator $H_{f}$ is self-adjoint with the known spectrum,
$[0,\infty)$,
and since $Q_j = Q_j -[0,\infty)$, we can write, using the spectral
theory,
\begin{equation} \label{V.13}
\mbox{l.h.s. of }\ \eqref{V.12}  = \sup_{\lambda \in Q_j } \|(H_{pg}
- \lambda)^{-1} {\overline P_{pj}} \|.
\end{equation}
Now, our claim follows from the definition \eqref{kappa_j} of
$\kappa_j$.

Since $\rho \le \kappa_{j}$, the inequalities \eqref{V.10} and
\eqref{V.12} imply
\begin{equation} \label{V.14}
\| R_0 (\lambda) \| \le 4 \rho^{-1}
\end{equation}
which implies \eqref{V.9} with $n=0$ and $C=4$.

The estimate \eqref{V.14}  and the relation $H_{0g}R_0(\lambda) =
\Ran \overline{\pi} + \lambda R_0(\lambda)$ imply the inequality
$\|H_{0g}R_0(\lambda)\| \le 2+4|e^{(p)}_0|/\rho$. Finally, since by
\eqref{Vbnd}, $\| |p|^2\psi\| \le 2\|H_{0g}\psi\|+ 2\|\psi\|$, we
have \eqref{V.9} with $n=1$.

The inequality \eqref{V.9} implies  the estimates
\begin{equation} \label{V.15}
\big\| \langle p\rangle^{2-n}(H_f
+1)^{n/2}(H_{0g}-\lambda)^{-1}\bar{\pi} \big\| \ \leq \ C \rho^{-1},
\end{equation}
for $n=1, 2$.
%

Now, we claim that
\begin{equation} \label{V.16}
\big\|  I_{g}(H_{0g}-\lambda)^{-1}\bar{\pi} \big\| \ \leq \ C g
\rho^{-1} \period
\end{equation}
Indeed, let $f(k)$ be an operator-valued function on $\cH_p$. Then
we have the following standard estimates
\begin{equation} \label{V.17}
\big\|  a(f) \psi \big\| \le \big(\int \frac{\|f
(k)\|_{part}^2}{|k|}d^3k \big)^\frac{1}{2}\big\|H_f^{1/2}\psi\big\|
\end{equation}
(cf. Eqn \eqref{VI.10} with $m+n =1$) and
\begin{equation} \label{V.18}
\big\|  a^*(f) \psi \big\|^2 = \int \|f (k)\|_{part}^2d^3k
\big\|\psi\big\|^2 +\big\|  a(f) \psi \big\|^2.
\end{equation}
Eqn \eqref{V.16} follows from the estimates Eqn \eqref{V.15},
\eqref{V.17} and \eqref{V.18}, the pull-through formula
\begin{equation} \label{pull-through}
a(k) \, f[\hf] \ = \ f[ \hf + |k| ] \, a(k),\
\end{equation}
and from the conditions on the operator $I_{g}$ imposed in Section
\ref{sec-IV}. For instance, for the term $W_{0,1}$ we have
$$\big\|W_{1,0} \psi \big\| \le \int_{|k| \le 1} \big\|w_{1,0}(k) \langle p\rangle^{-1} a^*(k)
\langle p\rangle\psi \big\|\frac{d^3k}{\sqrt{|k|}}
$$
$$\le \big(\int_{|k| \le 1}
\frac{\|w_{1,0}(k) \langle p\rangle^{-1} \|_{part}^2}{|k|}d^3k
\big)^\frac{1}{2}\big\|\langle p\rangle \psi\big\|$$
$$ + \big(\int_{|k| \le 1}
\frac{\|w_{1,0}(k) \langle p\rangle^{-1} \|_{part}^2}{|k|^2}d^3k
\big)^\frac{1}{2}\big\|H_f^{1/2}\langle p\rangle\psi\big\|$$
\begin{equation} \label{V.19}
 \le
\|w_{1,0} \|_{\mu}^{(0)} \big\|\langle
p\rangle(H_f+1)^{1/2}\psi\big\|
\end{equation}
for any $\mu > -1/2$. This together with  Eqn \eqref{V.9} implies
$\big\| W_{1,0}(H_{0g}-\lambda)^{-1}\bar{\pi} \big\| \ \leq \ C
\|w_{1,0} \|_{\mu}^{(0)}\ \rho^{-1}$. Now, the term $W_{0,2}$ is
estimated as follows:
$$\big\|W_{0,2} \psi \big\| \le \int_{|k_1| \le 1}\int_{|k_2| \le 1} \big\|w_{0,2}(k_1,k_2)  a(k_1)a(k_2)
\psi \big\|\frac{d^3k_1}{\sqrt{|k_1|}}\frac{d^3k_2}{\sqrt{|k_2|}}
$$
$$\le \int_{|k_1| \le 1}\big(\int_{|k_2| \le 1}
\frac{\|w_{0,2}(k_1,k_2)  \|_{part}^2}{|k_2|}d^3k_2
\big)^\frac{1}{2}\big\|H_f^{1/2}a(k_1)
\psi\big\|\frac{d^3k_1}{\sqrt{|k_1|}}$$
$$\le \int_{|k_1| \le 1}\big(\int_{|k_2| \le 1}
\frac{\|w_{0,2}(k_1,k_2)  \|_{part}^2}{|k_2|}d^3k_2
\big)^\frac{1}{2}\big\|(H_f +|k_1|)^{1/2}a(k_1)
\psi\big\|\frac{d^3k_1}{\sqrt{|k_1|}}$$
$$= \int_{|k_1| \le 1}\big(\int_{|k_2| \le 1}
\frac{\|w_{0,2}(k_1,k_2)  \|_{part}^2}{|k_2|}d^3k_2
\big)^\frac{1}{2}\big\|a(k_1)H_f^{1/2}
\psi\big\|\frac{d^3k_1}{\sqrt{|k_1|}}$$
$$\le \big(\int_{|k_1| \le 1}\int_{|k_2| \le 1}
\frac{\|w_{0,2}(k_1,k_2)  \|_{part}^2}{|k_1||k_2|}d^3k_1d^3k_2
\big)^\frac{1}{2}\big\|H_f \psi\big\|$$
\begin{equation} \label{V.19a}
 \le
\|w_{0,2} \|_{\mu}^{(0)} \big\|H_f\psi\big\|
\end{equation}
for any $\mu > -1$.
%

Eqn \eqref{V.16} implies that the series
%
%
%
%
\begin{equation} \label{V.20} \sum_{n=0}^{\infty}
(H_{0g} -\lambda)^{-1}\big(\overline{\pi} I_{g}R_0 (\lambda)\big)^n
\end{equation}
converges absolutely on the invariant subspace $\Ran
\overline{\pi}$, and is equal to $(H_{\tau_0 \overline{\pi}}$ $
-\lambda)^{-1}$, provided $g \ll \rho$. Estimating this series using
\eqref{V.16} gives the desired estimate \eqref{V.11a}. \qed

\textit{Proof of Theorem} \ref{thm-V.1}. The last proposition
together with the fact that the operators $\pi I_g$ and $I_g\pi$ are
bounded
yields Eqn \eqref{V.6}.
The second part of the theorem follows from the definition of the
Feshbach-Schur map, \eqref{Fesh}, the proposition and the Neumann
series argument. \qed

%


Note that $ K:= \cR_{\rho j}(H_g-\lambda)\mid _{\Ran
({\bar{P}}_{pj}\otimes\ \mathbf{1})}= (H_{0g}-\lambda)\mid_{ \Ran
({\bar{P}}_{pj}\otimes\ \mathbf{1})} $ and therefore $\sigma(K) =
\sigma(H_{pg})/\{\lambda_j\} +[0, \infty) - \lambda$. Hence for any
$\lambda \in Q_j$ we have
$$\min \{ | \mu - \lambda |\ \mid \mu \in \sigma (H_{pg})/\{\lambda_j
\} +[0,\infty)\} $$
\begin{equation}
\ge  \delta_j -|\lambda -\lambda_j| \ge \frac{1}{2}\delta_j.
\label{VI.22}
\end{equation}
Therefore $0 \notin \sigma(K)$.  This, the relation $\sigma
(\cR_{\rho j}(H_g-\lambda)) = \sigma(H_{\lambda j}) \cup \sigma(K)$
and Theorem \ref{thm-FeshSpec} of Appendix A imply
\begin{corollary} \label{cor-V.4} Let $\lambda \in Q_j.$ Then $\lambda \in \sigma(H_g)$
if and only if $0 \in \sigma(H_{\lambda j})$. Similar statement
holds also for point and essential spectra.
\end{corollary}
This corollary shows that to describe the spectrum of the operator
$H_g$ in the domain $Q_j$ it suffices to describe the spectrum of
the operators $H_{\lambda j}$ which act on the smaller space $\cF$.
In the next section we introduce a convenient Banach space which
contains the operators $H_{\lambda j}$ for $\lambda \in Q_j.$

Furthermore to prove bounds on resolvent in terms of bounds on
$H_{\lambda j}^{-1}$ one uses the relation
\begin{equation}
\cR_{\rho j}(H_g-\lambda)^{-1}=H_{\lambda j}^{-1}(P_{pj}\otimes\
\mathbf{1}) +(H_{0g}-\lambda)^{-1}({\bar{P}}_{pj}\otimes\
\mathbf{1}). \label{VI.27}
\end{equation}

\secct{A Banach Space of Hamiltonians} \label{sec-VI}
%
We construct a Banach space of Hamiltonians on which the
renormalization transformation will be defined. In order not to
complicate notation unnecessarily we will think about the creation-
and annihilation operators used below as scalar operators,
neglecting the helicity of photons.
We explain at the end of Supplement A how to reinterpret the
corresponding expression for the photon creation- and annihilation
operators.

%
%

Let $B_1^r$ denote the cartesian product of $r$ unit balls in
$\RR^{3}$, $I:=[0,1]$ and $m,n \ge 0$. Given functions $w_{m,n}:
I\times B_1^{m+n} \rightarrow \mathbb{C}, m+n > 0$, we consider
monomials, $W_{m,n} \equiv W_{m,n}[w_{m,n}]$, in the creation and
annihilation operators defined as follows:
%
%
%
%
\begin{eqnarray} \nonumber
&&W_{m,n}[w_{m,n}]  :=
\\ \label{VI.1}
&&  \int_{B_1^{m+n}} \frac{ dk_{(m,n)} }{ |k_{(m,n)}|^{1/2} } \;
a^*( k_{(m)} ) \, w_{m,n} \big[ \hf ; k_{(m,n)} \big] \, a(
\tk_{(n)} ) \:  .
\end{eqnarray}
%
Furthermore for $w_{0,0}: [0, \infty) \rightarrow \mathbb{C}$ we
define using the operator calculus $W_{0,0}:=w_{0,0}[H_f]$ (
$m=n=0$). Here we are using the notation
\begin{eqnarray} \label{VI.2}
& k_{(m)} \: := \: (k_1, \ldots, k_m) \: \in \: \RR^{3m} \comma
\hspace{5mm}
a^\#( k_{(m)} ) \: := \: \prod_{i=1}^m a^\#(k_i ),
\\  \label{VI.3}
& k_{(m,n)} \: := \: (k_{(m)}, \tk_{(n)}) \comma \hspace{5mm}
dk_{(m,n)} \: := \: \prod_{i=1}^m  d^3 k_i \; \prod_{i=1}^n d^3
\tk_i \comma &
\\  \label{VI.4}
& |k_{(m,n)}| \, := \, |k_{(m)}| \cdot |\tk_{(n)}| \comma
\hspace{3mm} |k_{(m)}| \, := \, |k_1| \cdots |k_m| , &
\end{eqnarray}
where $a^\#(k )$ stand for  $a(k )$ either or  $a^*(k )$. The
notation $W_{m,n}[w_{m,n}]$ stresses the dependence of $W_{m,n}$ on
$w_{m,n}$. Note that $W_{0,0}[w_{0,0}] := w_{0,0}[\hf]$.

We assume that, for every $m$ and $n$ with $m+n>0$ and for $s \ge
1$, the function $w_{m,n}[ r, , k_{(m,n)}]$
is $s$ times continuously differentiable in $r \in I$, for almost
every $k_{(m,n)} \in B_1^{m+n}$, and weakly differentiable in
$k_{(m,n)} \in B_1^{m+n}$, for almost every $r$ in $I$. As a
function of $k_{(m,n)}$, it is totally symmetric w.~r.~t.\ the
variables $k_{(m)} = (k_1, \ldots, k_m)$ and $\tk_{(n)} = (\tk_1,
\ldots, \tk_n)$ and obeys the norm bound
\begin{equation} \label{VI.5}
\| w_{m,n} \|_{\mu,s} \ :=
\sum_{n=0}^{s} \|  \partial_r^n w_{m,n} \|_{\mu} \ < \ \infty
\comma
\end{equation}
where $\mu \ge 0,\ s \ge 0$ and
%
%
%
\begin{equation} \label{VI.6}
\| w_{m,n} \|_{\mu} \ := \max_j \sup_{r \in I, k_{(m,n)} \in
B_1^{m+n}} \big| | k_j|^{-\mu}w_{m,n}[r ; k_{(m,n)}] \big|.
\end{equation}
Here and in what follows $k_j \in \mathbb{R}^3$ is the $j-$th
$3-$
vector in $k_{(m,n)}$ over which we take the supremum. For $m+n=0$
the variable $r$ ranges over $[0,\infty)$ and we assume that the
following norm is finite:
\begin{equation}
\\ \label{VI.7}
\ \| w_{0,0} \|_{\mu, s} := |w_{0,0}(0)|+ \sum_{1 \le n \leq s}
\sup_{r \in [0, \infty)}|
\partial_r^n w_{0,0}(r)|.
 \hspace{10mm}
\end{equation}
(This norm is independent of $\mu$, but we keep this index for
notational convenience.) The Banach space of functions $w_{m,n}$ of
this type is denoted by $\cW_{m,n}^{\mu,s}$.
%
%
%
%
%

We fix three numbers $\mu \ge0$, $0 < \xi < 1$ and $s \ge 1$ and
define the Banach space
\begin{equation} \label{VI.8}
\cW^{\mu,s} \ \equiv \cW^{\mu,s}_{\xi} := \ \bigoplus_{m+n \geq 0}
\cW_{m,n}^{\mu,s} \ \comma
\end{equation}
with the norm
\begin{equation} \label{VI.9}
\big\|  \uw \big\|_{\mu, s,\xi} \ := \ \sum_{m+n \geq 0}
\xi^{-(m+n)} \; \| w_{m,n} \|_{\mu, s} \ < \ \infty \period
\end{equation}
Clearly, $\cW^{\mu',s'}_{\xi'} \subset \cW^{\mu,s}_{\xi}$ if $\mu'
\ge \mu, s' \ge s$ and $\xi' \le \xi$.

Let $\chi_1(r) \equiv\chi_{r\le1}$ be a smooth cut-off function s.t.
$\chi_1 = 1$ for $r \le 9/10,\ = 0$ for $r\ge 1$ and $0 \le
\chi_1(r) \le1\ $  and $\sup|\partial^n_r \chi_1(r)| \le 30\ \forall
r$ and for $n=1,2.$ We denote $\chi_\rho(r) \equiv\chi_{r\le\rho}:=
\chi_1(r/\rho) \equiv\chi_{r/\rho\le1}$ and
$\chi_\rho\equiv\chi_{H_f\le\rho}$.

The following basic bound, proven in [2], links the norm defined in
\eqref{VI.6}
to the operator norm on $\cB[\cF]$.
%
\begin{theorem} \label{thm-VI.1}
Fix
$m,n \in \NN_0$ such that $m+n \geq 1$. Suppose that $w_{m,n} \in
\cW_{m,n}^{0,1}$, and let $W_{m,n} \equiv W_{m,n}[w_{m,n}]$ be as
defined in (\ref{VI.1}). Then for all $ \lambda >0$
%
\begin{equation} \label{VI.10}
\big\|  (\hf+\lambda)^{-m/2} \, W_{m,n} \, (\hf+\lambda)^{-n/2}
\big\| \ \leq \  \, \| w_{m,n} \|_{0} \, ,
\end{equation}
and therefore
\begin{equation} \label{VI.11}
\big\| \chi_\rho \, W_{m,n} \, \chi_\rho
 \big\|
\ \leq \ \frac{\rho^{(m+n)(1+\mu)}}{\sqrt{m! \, n!} } \, \| w_{m,n}
\|_{0} \, ,
\end{equation}
where $\| \, \cdot \, \|$ denotes the operator norm on $\cB[\cF]$.
\end{theorem}

Theorem~\ref{thm-VI.1} says that the finiteness of $\| w_{m,n}
\|_{0}$ insures that $W_{m,n}$ defines a bounded operator on
$\cB[\cF]$.

With a sequence $\uw := (w_{m,n})_{m+n \geq 0}$ in $\cW^{\mu,s}$ we
associate an operator
by setting
\begin{equation} \label{VI.12}
H(\uw)  := W_{0,0}[\uw] + \sum_{m+n \geq 1} \chi_1W_{m,n}[\uw]\chi_1
\end{equation}
where we write $W_{m,n}[\uw] := W_{m,n}[w_{m,n}]$. The r.h.s. of
\eqref{VI.12} are said to be in \textit{generalized normal (or
Wick-ordered) form} of the operator $H(\uw)$. Theorem~\ref{thm-VI.1}
shows that the series in (\ref{VI.12}) converges in the operator
norm and obeys the
estimate
\begin{equation} \label{eq-III-1-25.1}
\big\| \, H(\uw)- W_{0,0}(\uw) \, \big\|
\leq \ \xi\big\| \,
\uw_1 \, \big\|_{\mu,0, \xi} \comma
\end{equation}
for arbitrary $\uw = (w_{m,n})_{m+n \geq 0} \in \cW^{\mu,0}$ and any
$\mu
> -1/2$. Here $\uw_1 = (w_{m,n})_{m+n \geq 1}$. Hence the
linear map
\begin{equation} \label{eq-III-1-24.1}
H : \uw \to H(\uw)
\end{equation}
takes $\cW^{\mu,0}$ into the set of closed operators on Fock space
$\cF$.
The following result is proven in [2].
%
\begin{theorem} \label{thm-III-1-2}
For any $\mu \ge 0$ and $0 < \xi < 1$, the map $H : \uw \to H(\uw)$,
given in (\ref{VI.12}), is injective.
%
%
\end{theorem}

Furthermore, we
define the Banach space
%
%
%
%
%
%
\begin{equation} \label{eq-III-1-17}
\cW_{1}^{\mu,s} \ := \ \bigoplus_{m+n \geq 1} \cW_{m,n}^{\mu,s}
\comma
\end{equation}
to be the set of all sequences $\uw_1 := (w_{m,n})_{m+n \geq 1}$
obeying
\begin{equation} \label{VI.17}
\| \uw_1 \|_{\mu, s,\xi} \ := \ \sum_{m+n \geq 1} \xi^{-(m+n)} \; \|
w_{m,n} \|_{\mu,s}\ < \ \infty \period
\end{equation}

We define the spaces $\cW_{op}^{\mu,s} :=H(\cW^{\mu,s})$,
$\cW_{1,op}^{\mu,s} :=H(\cW_1^{\mu,s})$ and $\cW_{mn,op}^{\mu,s}
:=H(\cW_{mn}^{\mu,s})$. Sometimes we display the parameter $\xi$ as
in $\cW_{op,\xi}^{\mu,s} :=H(\cW^{\mu,s}_\xi)$. Theorem
\ref{thm-III-1-2} implies that $\cW_{op}^{\mu,s} :=H(\cW^{\mu,s})$
is a Banach space under the norm $\big\| \, H(\uw) \big\|_{\mu,s,
\xi}$ $:=\ \big\| \, \uw \, \big\|_{\mu,s, \xi}$.  Similarly, the
spaces $\cW_{1,op}^{\mu,s}$ and $\cW_{mn,op}^{\mu,s}$ are also
Banach spaces in the corresponding norms.

In this paper we need and consider only the case $s=1$. However, we
keep the more general notation for convenience of references
elsewhere.

\secct{The operator $\cR_{\rho j}(H_g -\lambda)$}
\label{sec-VII}

In this section we give a detailed description of the family of
operators $H_{\lambda j}:=\cR_{\rho j}(H_g -\lambda) \mid _{\Ran
(P_{pj}\otimes\ \textbf{1})}$ (see Theorem \ref{thm-V.1}). Here,
recall, that $P_{pj}$ denotes the projection on the particle
eigenspace corresponding to the eigenvalue $\lambda_j$. We define
the following polydisc in $\cW_{op}^{\mu,s}$:
\begin{eqnarray} \label{disc}
& & \cD^{\mu,s}(\alpha,\beta,\gamma) :=  \Big\{ H(\uw) \in
\cW_{op}^{\mu,s} \ \Big| \
 |w_{0,0}(0)| \leq \alpha \comma
\\ \nonumber & &
 \sup_{r \in [0,\infty)}|\partial_r w_{0,0}(r) - 1 | \leq \beta
 \comma \hspace{4mm}
\| \uw_1 \|_{\mu,s, \xi} \leq \gamma \Big\} \comma
\end{eqnarray}
for $\alpha, \beta, \gamma >0$. Recall that $\uw_1 := (w_{m,n})_{m+n
\geq 1}$.
In what follows we fix the parameter $\xi$ in \eqref{disc} as
$\xi=1/4 $.
%

%
\begin{theorem}
\label{thm-Hlambdaj} Let $H_g$ be a Hamiltonian of the class
$GH_\mu$ defined in Section \ref{sec-IV}
%
with $ \mu \ge 0$. We assume that $\delta_j > 0$.
%
Then for $g \ll \rho\le min(\kappa_{j}, 1/2)$ and $\lambda\in Q_j$,
\begin{equation}
H_{\lambda j}-\rho^{-1}(\lambda_{j}-\lambda)\in
\cD^{\mu,s}(\alpha,\beta,\gamma), \label{Hlambdaj}
\end{equation}
where $\alpha=O(g^2 \rho^{\mu -2})$, $\beta=O(g^2\rho^{\mu -1})$,
$\gamma =O(g\rho^\mu)$.
\end{theorem}

Note that if $\psi_j^{(p)}$ is an eigenfunction of $H_{pg}$ with the
eigenvalue $\lambda_j$
and $\Psi_j:=\psi_j^{(p)}\otimes\Omega$, then we have
\begin{equation*}
\lambda_{j}-\lambda=\la H_g-\lambda\ra_{\Psi_j}.
\end{equation*}

The proof of
Theorem~\ref{thm-Hlambdaj} follows the lines of the proof of Theorem
IV.3
of \cite{FroehlichGriesemerSigal2008b}. It is similar to the proofs
of related results of
\cite{BachFroehlichSigal1998a,BachFroehlichSigal1998b,BachFroehlichSigal1999}.
However, there are a few differences here. The main ones are that we
have to deal with unbounded interactions and, more importantly, with
momentum-anisotropic spaces. 
Since the proof of Theorem~\ref{thm-Hlambdaj} is technically rather
involved, it is delegated to an appendix, Appendix B.
%
\secct{Spectrum of $H_g$} \label{sec-VIII}
In this section we describe the spectrum of the operator $H_g \in
GH_\mu$ defined in Section \ref{sec-IV}. We begin with some
definitions. Recall that $D(\lambda,r):=\{z \in \mathbb{C} |
|z-\lambda| \le r \}$, a disc in the complex plane. Denote
$\cD:=\cD^{\mu,1}(\alpha,\beta,\gamma)$ with $\alpha, \beta,\gamma
\ll 1$ and let $\cD_s:=\cD^{\mu,1}(0,\beta,\gamma)$ (the subindex s
stands for 'stable', not to be confused with the smoothness index
$s$ which in this section is taken to be 1). For $H \in \cD$ we
denote $H_u:=\la H\ra_\Omega$ and $H_s:=H - \la H\ra_\Omega\
\mathbf{1}$ (the unstable- and stable-central-space components of
$H$, respectively). Note that if  $H \in \cD$, then $H_s \in \cD_s$.

Recall that a complex function $f$ from an open set $\cD$ in a
complex Banach space $\cB$ 
is said to be \textit{analytic} iff $\forall H \in \cD$ and $\ \forall \xi \in \cB,\ f(H+ \tau \xi)$ is analytic in the complex variable $\tau$ for $|\tau|$ sufficiently small (see \cite{Berger}). (One can show that $f$ is analytic iff it is 
G\^{a}teaux-differentiable (\cite{Berger,
HillePhillips}). A stronger notion of analyticity, requiring in addition that $f$ is locally bounded, is used in \cite{HillePhillips}.) Furthermore, if $f$ is analytic in $\cD$ and $g$ is an analytic vector-function from an open set $\Omega$ in
$\mathbb{C}$ into  $\cD$, then the composite function $f\circ g$ is
analytic on $\Omega$. In what follows $\cB$ is the space 
of $H_f$-bounded operators on $\cF$.

%

Our analysis uses the following result from
\cite{FroehlichGriesemerSigal2008b}:
\begin{theorem} \label{thm-Hspec}
For $\alpha, \beta$ and $\gamma$ sufficiently small there is
an analytic  map $e:\cD_s \rightarrow D(0, 4\alpha)$ s.t. $e(H) \in
\mathbb{R}$ for $H=H^*$ and for any $H \in \cD_s$,
$\sigma(H) \subset e(H) +S$, where
\begin{equation}S:=\{w\in \mathbb{C}| \rRe w \ge 0, |\rIm
w| \le \frac{1}{3} \rRe w \}. \label{S}
\end{equation}
Moreover, the number $e(H)$ is
an eigenvalue of the operator $H$.
\end{theorem}

Let $H_g $ be in the class $ GH_\mu$ defined in Section \ref{sec-IV}
with $\mu >0$ and let $H_{z j}$ be  the operator obtained from $H_g
$ according to Theorem \ref{thm-V.1}.   By Corollary \ref{cor-V.4},
for $z \in Q_j,$ we have that $z \in \sigma(H_g)$ if and only if $0
\in \sigma(H_{z j})$ and similarly for point and essential spectra.
By Theorem \ref{thm-Hlambdaj}, $\forall z \in Q_j,\ H_{z j} \in
\cD^{\mu, 1} (\alpha, \beta, \delta)$ with $\alpha=O(g^2
\rho^{-1}),$ $ \beta=O(g^2)$ and $\gamma=O(g\rho^\mu)$.
Since by our assumption $g \ll 1$, we can choose $\rho$ (under the
restriction $g \ll \rho\le \min(\kappa_j, 1/2)$) so that
\begin{equation} \label{eqn:32}
g^2\rho^{-1},\ g\rho^\mu\ll 1.
\end{equation}
In this case the condition
of Theorem \ref{thm-Hspec} is satisfied for $H_{zjs} \in \cD_s$.
Therefore it is in the domain of the map $e:\cD_s\rightarrow
\mathbb{C}$ described in Theorem \ref{thm-Hspec} above and we can
define
\begin{equation} \label{phi}
\varphi_j (z) : = E_j (z) + e (H_{zj s}),
\end{equation}
where $E_j (z) : = H_{zj u} = \langle \Omega, H_{zj } \Omega
\rangle$ and $z \in Q_j$. Let $\Gamma_\rho$ be the unitary
dilatation on $\cF$ defined by
\begin{equation} \label{eq-III-2-1a}
\Gamma_\rho = U_f(-\ln\rho)
\end{equation}
%
where $U_f(-\ln\rho)$ is defined in Section \ref{sec-III}. Our goal
is to prove the following

\begin{theorem} \label{thm-specHg}
Let the Hamiltonian $H_g $ be in the class $ GH_\mu$ defined in
Section \ref{sec-IV} with $\mu >0$ and let $g \ll \kappa_j$. Then:

(i) The equation $\varphi_j (\epsilon) = 0$ has a unique solution
$e_j \in Q_j$ and this solution obeys the estimate $|e_j-\lambda_j|
\le 15 \alpha$;

(ii)
$e_j$ is an eigenvalue of $H_{g }$ and
\begin{align} \label{VI.25}
\sigma (H_{g }) &\cap Q_j \subset \{ z \in Q_j \mid \rRe
(z - e_j)\ge 0\ \\
\nonumber & \hbox{and}\ | \rIm (z - e_j) | \le \frac{1}{2} | \rRe (z
- e_j) | \} ;
\end{align}

(iii)
If $\psi_{ j}$ is an eigenfunction of the operator $H_{e_{j} j}$
corresponding to the eigenvalue $0$, then the vector
\begin{equation} \label{eq-III-4-28.2}
\Psi_j :=Q_{ \pi} \big( \, H_g -e_{j} \big)\Gamma_{\rho}^*\psi_{j}
\neq 0,
\end{equation}
%
where $\pi$ and $Q_{\pi} \big( H \big)$ are defined in Eqns
~\eqref{pi} and \eqref{Qchi}, respectively, is an eigenfunction of
the operator $H_g$
corresponding to the eigenvalue $e_{j}$.
\end{theorem}
\Proof In this proof we omit the subindex $j$. (i) Since $e:\cD_s
\rightarrow D(0, 4\alpha)$  is an analytic map, $z \rightarrow
H_{zs}$  is an analytic vector-function and $z \rightarrow E(z)$ is
an analytic function on  $Q^{int}$, by Theorem \ref{thm-V.1}, we
conclude that the function $\varphi$ is analytic on $Q^{int}$. Here
$Q^{int}$ is the interior of the set $Q$.

Furthermore, the definitions \eqref{phi} and  $\Delta_0 E(z) := E
(z) -\rho^{-1}(\lambda - z)$ (remember that in this proof $\lambda =
\lambda_j$) imply that $\varphi(\lambda)=\Delta_0 E(\lambda)+ e
(H_{\lambda s}).$

We have, by Theorem \ref{thm-Hlambdaj}, that $|\Delta_0 E(\lambda)|
\leq \alpha$. Hence $|\varphi(\lambda)| \le 5\alpha.$
Furthermore since $Q$ is inside a square in $\mathbb{C}$ of side
$\delta/3$, we have, by the Cauchy formula, that
\begin{equation} \label{VIII.7}
|\partial_{z}^m \Delta_0 E(z)|\le\alpha (3/ \delta)^{m}\ for\  m=
0,1.
\end{equation}
(remember that in this proof $\delta = \delta_j$). Similarly we
have:
\begin{equation} \label{VIII.8}
|\partial_z e(H(z)_s)| \le 4\alpha (3/ \delta)^{-1}.
\end{equation}
The last two inequalities and the equation $\Delta_0 E(z) := E (z)
-\rho^{-1}(\lambda - z)$ give
\begin{equation} \label{VIII.9}
|\partial_z \varphi (z) +1| \le 15\alpha / \delta.
\end{equation}
Hence by inverse function theorem, for $\alpha$ sufficiently small
the equation $\varphi (z) =0$ has a unique solution, $e$, in $Q$ and
this solution satisfies the bound $|e-\lambda| \le 15 \alpha$.

(ii) By Theorem \ref{thm-Hspec}, $\varphi (z)$ is an eigenvalue of
the operator $H_{z} =E(z)+ H_{zs}$. Hence $0$ is  an eigenvalue of
the operator $H_{e}$. By Corollary \ref{cor-V.4}, $z$ is an
eigenvalue of  $H_{g } \leftrightarrow 0$ is an eigenvalue of
$H_{z}.$ Hence $e$ is  an eigenvalue of the operator $H_{g}$.

Next, by Corollary \ref{cor-V.4}, we have for any $z \in Q$
\begin{equation} \label{VI.26}
z \in \sigma (H_{g }) \leftrightarrow 0 \in \sigma  (H_{z}).
\end{equation}
Due to Theorem \ref{thm-Hspec} we have that
$\sigma(H_{z})=E(z)+\sigma(H_{zs})\subset E(z)+e(H_{zs})+S =
\varphi(z)+S$, where the set $S$ is defined in \eqref{S}.  This
together with \eqref{VI.26} gives $z \in \sigma (H_{g })\cap Q
\leftrightarrow \varphi(z)\in -S$ or
\begin{equation}
\sigma  (H_{g })\cap Q = \varphi^{-1} (-S).
\end{equation}

Now the second part of the proof will follow if we show that
$\varphi^{-1} (-S)$ is a subset of the r.h.s. of \eqref{VI.25}.
Denote $\mu : = z - e$ and let
\begin{equation} \label{Im1}
| \rIm \mu | > {1 \over 2} | \rRe \mu |.
\end{equation}
Let $w : = -\varphi (z)$.  
%
Using that $\varphi (e) = 0$  and the integral of derivative formula we find
\begin{equation} 
\varphi (z) = (z - e)g(z)
\end{equation}
with $g(z):= \int_0^1 \varphi (e + s(z -e))ds$.
This gives
\begin{equation} \label{Im2}
| \rIm w | = | \rRe  g \rIm \mu + \rIm g \rRe \mu |.
\end{equation}
Now, the definitions \eqref{phi} and  $\Delta_0 E(z) := E (z)
-\rho^{-1}(\lambda - z)$ (remember that in this proof $\lambda =
\lambda_j$) imply that
\begin{equation} \label{VI.31}
\partial_z \varphi (z) = -1 + \partial_z \Delta_0
E(z) + \partial_z e (H_{z s}).
\end{equation}
%
%
%
%
%
%
%
%
This, the fact that $ \bar{z}: =e + s(z -e) \in Q$ and Eqns \eqref{VIII.7} and \eqref{VIII.8} give 
\begin{equation} \label{ReIm}
| \rRe  g (\overline z) | \ge 1 - O (\alpha)\ \hbox{and}\ |
\rIm g (\overline z) | \le O (\alpha).
\end{equation}
Relations \eqref{Im2} and \eqref{ReIm} imply the estimate
\begin{equation*}
| \rIm w | \ge (1 - O (\alpha)) | \rIm \mu | - O (\alpha) | \rRe \mu
|
\end{equation*}
which together with \eqref{Im1} gives
\begin{equation} \label{VI.33}
|\rIm w | \ge {1 \over 4} (1 - O (\alpha)) | \rIm \mu | + {3 \over
8} (1 - O (\alpha)) | \rRe \mu |.
\end{equation}
Similarly, we obtain
\begin{align}
| \rRe w | &= | \rRe  g \rRe \mu - \rIm g \rIm \mu |
\\ \nonumber &\le (1 + O (\alpha)) | \rRe \mu | + O (\alpha) | \rIm \mu |.
\end{align}
The last two relations imply $| \rIm w | > {1 \over 3} | \rRe w |$
and therefore $w \not\in S$ or what is the same $z \not\in
\varphi^{-1} (-S)$.

Now let $\rRe \mu < 0$. Then Eqns \eqref{VI.31}-\eqref{ReIm} imply
that $\rRe w = - \rRe  g \rRe \mu$ $ + \rIm g \rIm \mu
= (-1+O(\alpha))|\rRe \mu|+O(\alpha \rIm \mu)$. Thus, $\rRe w =
(-1+O(\alpha))|\rRe \mu|$, provided $|\rIm \mu| \le |\rRe \mu|$.
Hence also in this case we have  $z \not\in \varphi^{-1} (-S)$. Thus
we conclude that $\varphi^{-1} (-S)$ is a subset of the set on the
r.h.s. of \eqref{VI.25}, as claimed.

(iii) Finally, the last part of the theorem follows from
Theorem~\ref{thm-FeshSpec}(iii) of Appendix A.
Theorem \ref{thm-specHg} is proven. \qed

%
%
%

\secct{Proof of Theorems \ref{thm-main} and \ref{thm-main2}}
\label{sec-IX}


We begin with the proof of existence of the ground state. Let $H_g$
be a Hamiltonian from the class $GH_\mu,\ \mu >0$ defined in Section
\ref{sec-IV}. We assume that $H_g$ is self-adjoint. Special cases of
such Hamiltonians $H_g$ are the Pauli-Fierz and Nelson Hamiltonians,
$H^{PF}_g$ and $H^{N}_g$, given in \eqref{Hpf} and \eqref{Hn},
respectively.
Then the operator $H_g,\ g \ll  \kappa_{0},$ clearly satisfies the
conditions of Theorem \ref{thm-specHg} with $j=0$. Moreover, the
particle Hamiltonian $H_{pg}$ entering $H_{g}$ is self-adjoint which
implies that the constant $\kappa_{0}$, defined in  \eqref{kappa_j},
is $\kappa_{0} = \dist (\sigma (H_{pg} |_ {Ran \overline P_{p0}}),
Q_j) \ge \delta_0 /2$. Here, recall, $\delta_0:=\dist(\lambda^{}_0,
\sigma(H_{pg})/\{\lambda^{}_0\})$, where $\lambda^{}_0$ is the
smallest eigenvalues of the operator $H_{pg}$. This implies the
existence of the ground state for $H_g,\ g \ll \delta_0$.

Now, $H^{PF}_p = H_p+ O(g^2)$ (see \eqref{Hpg}) and $H^{N}_p = H_p+
O(g^2)$ (see the paragraph after \eqref{H_0}). Hence, if $H_g$ is
either $H^{PF}_g$ or $H^{N}_g$, then $\lambda^{}_j
=\epsilon^{(p)}_{j}+O(g^2)$ and $\delta_0
=\epsilon^{(p)}_{gap}(\epsilon^{(p)}_0)+O(g^2)$, where, recall,
where $\lambda^{}_j$ are the eigenvalues of the operator $H_{pg}$
labeled in order of their increase and counting their
multiplicities, $\epsilon^{(p)}_{j}$ are the eigenvalues of the
operator $H_p$ given in \eqref{Hp} and
$$\epsilon^{(p)}_{gap}(\nu):=\min \{|\epsilon^{(p)}_i -\epsilon^{(p)}_j |
 |i\neq j,\ \epsilon^{(p)}_i, \epsilon^{(p)}_j \le \nu \}.$$
Consequently, it suffices to assume that $g \ll
\epsilon^{(p)}_{gap}(\epsilon^{(p)}_0)$. Since  $H^{PF}_g$is unitary
equivalent to  $H^{SM}_g$, this proves the part of the statement of
Theorem \ref{thm-main} concerning the ground state.

Note that  the energy of the found ground state solves the equation
$\varphi_0(\epsilon) = 0$ (see \eqref{phi} for the definition of
$\varphi_j(\epsilon)$).

Now we prove Theorem \ref{thm-specHtheta} which implies the part of
the statement of Theorem \ref{thm-main} concerning the excited
states and  Theorem \ref{thm-main2}. Let $H_g :=
e^{\theta}H^{\#}_{g\theta}$ where $H^{\#}_{g\theta}$ is the complex
deformation of the Hamiltonian $H^{\#}_{g}$, which is either the
Pauli-Fierz Hamiltonian, $H^{PF}_g$,
or the Nelson Hamiltonian, $H^{N}_g$,
defined in \eqref{III.5} and in \eqref{XI.5}, respectively. Then the
Hamiltonian $H_g$ belongs to the class $GH_\mu$ defined in Section
\ref{sec-IV} with $\mu>0$. We will assume $0 < \rIm \theta \le
\min(\theta_0, \pi/4)$, where $\theta_0$ is defined in Condition
(DA) in Section \ref{sec-I}, and $\rRe \theta = 0$ and we will
assume  $g \ll \min(\kappa_{j}, \epsilon^{(p)}_{gap}(\nu))$.
%

Let $H^{\#}_{p}$ and $H^{\#}_{p\theta}$  be the particle
Hamiltonians entering $H^{\#}_{g }$ and $H^{\#}_{g \theta}$,
respectively. We show that $$\delta_j = \dist (\lambda_j,
\sigma(H_{pg})/\{\lambda_j\}+ \overline{\mathbb{R}^+}) >0$$  for the
particle Hamiltonian $H_{pg}:=e^{\theta}H^{\#}_{p\theta}$, entering
$H_{g}$, provided $j \le j(\nu)$, with $\nu < \inf
\sigma_{ess}(H_p^{})$, and $g \ll \epsilon^{(p)}_{gap}(\nu)$. Here,
recall, $\lambda_j$ are the eigenvalues of the operator
$H_{pg}:=e^{\theta}H^{\#}_{p\theta}$, $j(\nu):= \max\{j:
\epsilon^{(p)}_{j} \le \nu\}$ and $\epsilon^{(p)}_{gap}(\nu)$ is
defined above. To do this we note first that, since $H_p^{\#} =
H_p^{} +O(g^2)$, we have $\nu < \inf \sigma_{ess}(H_p^{\#})$ for $g$
sufficiently small. Furthermore, since we have chosen
$\rRe\theta=0$, we have that $\delta_j = \dist (\epsilon^{\#}_j,
\sigma(H^{\#}_{p\theta})/\{\epsilon^{\#}_j\}+ e^{-\theta}
\overline{\mathbb{R}^+})$, where $\epsilon^{\#}_i=
e^{-\theta}\lambda_i$ are eigenvalues of the operator
$H^{\#}_{p\theta}$. By the definition of the operator
$H^{\#}_{p\theta}= -\frac{1}{2}e^{-2\theta}\Delta+ V_{g\theta}$ and
the Balslev-Combes-Simon theorem (remember that $V_{g\theta}=
V_{\theta} +O(g^2)$ and that $ V_{\theta}$ is $\Delta-$compact, by
Condition (V), which implies the $\Delta-$compactness of $V$ in the
one particle case, and Condition (DA), which implies the
$\Delta-$compactness of $V_{\theta}$ in the one particle case, of
Section \ref{sec-I})
we have that it has no complex eigenvalues in the domain $\{\rRe z
\le \nu\}$ and therefore its eigenvalues $\epsilon^{\#}_j,\ j \le
j(\nu),$ coincide with the eigenvalues of the operator $H^{\#}_{p}$
which are $ \le \nu$. Hence we have that $$\delta_j =  \min (\dist
(\epsilon^{\#}_j, \sigma(H^{\#}_{p})/\{\epsilon^{\#}_j\}),
(\epsilon^{\#}_{j-1} -\epsilon^{\#}_j)\tan(\rIm\theta))$$ and
therefore $\delta_j
>0$.

Thus, for any $j \le j(\nu)$, the operator $H_g
(:=e^{\theta}H^{\#}_{g\theta}),\ g \ll \min(\kappa_{j},
\epsilon^{(p)}_{gap}(\nu)),$ satisfies the conditions of Theorem
\ref{thm-specHg}. This implies that the spectrum of $H^{\#}_{g
\theta}$ near $\epsilon_j=e^{-\theta} \lambda_j$ is of the form
\begin{align} \label{VII.1}
\sigma (H^{\#}_{g \theta}) &\cap e^{-\theta}Q_j \subset \{ z \in
e^{-\theta}Q_j \mid \rRe
(e^{\theta} (z - \epsilon_j))\ge 0\ \\
\nonumber & \hbox{and}\ \mid \rIm (e^{\theta} (z - \epsilon_j)) \mid
\le \frac{1}{2} \mid \rRe (e^{\theta} (z - \epsilon_j)) \mid \} ,
\end{align}
where $\epsilon_j \in e^{-\theta}Q_j$ is an eigenvalue of $H^{\#}_{g
\theta}$. Moreover, $e^{\theta}\epsilon_j$ is the unique solution to
the equation $\varphi_j (\epsilon) = 0$ and $\epsilon_j \rightarrow
\epsilon^{\#}_j$ as $g \rightarrow 0$.

Let $\varphi_j(\epsilon, \theta)\equiv \varphi_j(\epsilon)$ be the
function constructed in \eqref{phi} for the operator $H_g
:=e^{\theta}H^{\#}_{g\theta}$. It is not hard to see that
$\varphi_j(\epsilon, \theta)$ is analytic in $\theta$. Since by
Theorem \ref{thm-specHg} $e^{\theta}\epsilon_j$   is a unique
solution to the equation $\varphi_j(\epsilon, \theta)=0$ we conclude
that $\epsilon_j$  is analytic in (a fractional power of) $\theta$.
On the other hand, by Eqn \eqref{III.4},  $\epsilon_j$  is
independent of $\rRe \theta$. Hence it is independent of $\theta$.

The eigenvalue $\epsilon_0$ is always real and therefore is the
eigenvalue also of $H_g^{\#}$. This is the ground state energy of
$H_g^{\#}$. For $j>0$ the  eigenvalue $\epsilon_j$ can be either
complex or real, i.e. either a resonance or an eigenvalue of
$H_g^{\#}$. (If the (FGR) condition is satisfied then $\rIm
\epsilon_j < 0$ for $j \ne 0$ and, in fact, $\rIm \epsilon_j = -
\gamma_j g^2 +O(g^4)$ for some $\gamma_j >0$ independent of $g$, see
\cite{BachFroehlichSigal1999}). In the degenerate case, the total
multiplicity of the resonances and eigenvalues arising from
$\epsilon^{\#}_j$ is equal to the multiplicity of $\epsilon^{\#}_j$.

Thus we have proven all the statements of Theorem
\ref{thm-specHtheta}, but under the stronger assumption $g \ll
\min(\kappa_{j}, \epsilon^{(p)}_{gap}(\nu))$. Now we relax this
assumption.

Define $\delta^{\#}_j := \dist(\epsilon^{\#}_j,
\sigma(H^{\#}_{p})/\{\epsilon^{\#}_j\})$. The following proposition
states that the restrictions $g \ll \delta^{\#}_j$ and $|\rIm \theta
| \ll \delta^{\#}_{j}$ imply the restriction $g \ll \kappa_j$.
Recall that $\kappa_{j}$ and $\delta_{j}$ are defined in Eqns
\eqref{kappa_j} and \eqref{delta_j}, respectively.
\begin{proposition} \label{prop-V.4} Assume that $|\rIm \theta | \ll \delta^{\#}_{j}$. Then
there is a numerical constant $c >0$ s.t.
$\kappa_{j} \ge c\delta^{\#}_{j} \tan(\rIm \theta)$.
\end{proposition}
\Proof Observe first that this proposition concerns entirely the
particle Hamiltonian $H_{pg }:=e^\theta H^{\#}_{p \theta}$. In its
proof we omit the subindices $p$ and $g$.

First we estimate $\delta_j$ in terms of $\delta^{\#}_j
.$ We assume $\rRe \theta =0$. By the definitions of $\delta_j$ and
of $H:=e^\theta H^{\#}_{ \theta}$ we have $\delta_j=
\dist(\epsilon^{\#}_j,
\sigma(H^{\#}_{\theta})/\{\epsilon^{\#}_j\} +e^{-\theta}
\overline{\mathbb{R}^+}).$ Since
$\sigma(H^{\#}_{\theta})=\{\epsilon^{\#}_i\}\bigcup
e^{-2\theta}\overline{\mathbb{R}^+}$, this gives
$$\delta_j=\min[\dist(\epsilon^{\#}_j,
\sigma(H^{\#})/\{\epsilon^{\#}_j\}), \dist(
\epsilon^{\#}_j,\epsilon^{\#}_{j-1}+
e^{-\theta}\overline{\mathbb{R}^+})]$$ which can be rewritten as
\begin{equation} \label{VI.2}
\delta_j= \min(\delta^{\#}_j, (\epsilon^{\#}_j-
\epsilon^{\#}_{j-1})\tan (\rIm \theta)).
\end{equation} This, in particular, gives
$
\delta^{\#}_j \ge
\delta_j \ge \delta^{\#}_j
\tan (\rIm \theta).
$

Now we estimate the norm on the r.h.s. of Eqn \eqref{kappa_j}. We
begin with the case of $\delta=0$. In what follows $\lambda \in Q_j$
is fixed. First, we write ${\overline P_{j}} = P_{<j}+P_{>j}$, where
$P_{<j} := \sum_{i < j}P_{i}$ and $P_{>j} := \mathbf{1} - \sum_{_{i
\le j} }P_{i}$. Here, recall, $P_i$ are the eigenprojections of $H_{
}:=e^\theta H^{\#}_{\theta}$ corresponding to the eigenvalues
$\lambda_i$. Since $(H^{}_{} - \lambda)^{-1} P_{<j} = \sum_{i <
j}(\lambda_{i} - \lambda)^{-1}P_{i}$, we have $\|(H^{}_{} -
\lambda)^{-1} P_{<j} \| \le
%
%
C(\min_{i <j}|\lambda_i - \lambda|)^{-1}.$ To estimate the r.h.s. of
the above inequality we write for $\lambda \in Q_j$
$$\min_{i <j}|\lambda_i-\lambda|\ge \min_{i
<j}|\rIm(\lambda_i-\lambda)|$$ $$\ge \min_{i
<j}|\rIm(\lambda_i-\lambda_j)|-|\rIm(\lambda_j-\lambda)|. $$ By the
definitions of $\delta_j$ and $Q_j$ (see Eqns \eqref{delta_j} and
\eqref{Q_j}) and by Eqn \eqref{VI.2}, we have
$|\rIm(\lambda_j-\lambda)| \le \frac{1}{3}\delta_j \le
\frac{1}{3}(\epsilon^{\#}_j- \epsilon^{\#}_{j-1})\tan (\rIm
\theta))$. On the other hand, $|\rIm(\lambda_i-\lambda_j)|=
(\epsilon^{\#}_j- \epsilon^{\#}_{i})\sin(\rIm \theta)$. Hence
$$\min_{i <j}|\lambda_i-\lambda| \ge
(\epsilon^{\#}_j- \epsilon^{\#}_{j-1})(\sin(\rIm \theta)
-\frac{1}{3}\tan (\rIm \theta)).
%
%
%
$$ For $0<\rIm \theta \le \pi/3$, this gives $\min_{i
<j}|\lambda_i-\lambda| \ge \frac{1}{3}\delta^{\#}_j \sin (\rIm
\theta)
$ for any $\lambda \in Q_j$. This, together
with the estimate derived above, yields
\begin{equation} \label{IX.3}
\|(H^{}_{} - \lambda)^{-1} P_{<j} \| \le C({\delta^{\#}_j}\sin (\rIm
\theta))^{-1}.
\end{equation}

To estimate $(H^{}_{} - \lambda)^{-1}  P_{>j}$
%
%
we write it as the contour integral
%
%
\begin{equation} \label{ContourInt2}
(H^{}_{} - \lambda)^{-1} P_{>j}= \frac{1}{2\pi
i}e^{-\theta}\oint_\Gamma \big( H^{\#}_{\theta}-z \big)^{-1} (z-
e^{-\theta}\lambda)^{-1}dz,
\end{equation}
where the contour $\Gamma$ is defined as
%
%
$\Gamma := \mu + i\mathbb{R}$, where $\mu :=
\frac{1}{4}\epsilon^{\#}_j + \frac{3}{4}\epsilon^{\#}_{j+1}$.
%
%
%
%

Next, expanding $e^{2\theta}V_g(e^{\theta}x)$ in $\theta$, we have
$H^{\#}_{\theta} = e^{-2\theta}H^{\#}_{} +O(\theta)$. Hence for
$|\rIm \theta | \ll \inf_{z\in \Gamma}\dist(z,
\sigma(H^{\#}_{\theta}))$ and $\rRe \theta =0$, this gives
$$\|(H^{\#}_{\theta} - z)^{-1} \| \le 2 \|(e^{-2\theta}H^{\#}_{} - z)^{-1} \| \le
2/\dist(z,\sigma(e^{-2\theta}H^{\#}_{})).$$ Again, by
$H^{\#}_{\theta} = e^{-2\theta}H^{\#}_{} +O(\theta)$ and the
condition $|\theta| \ll \inf_{z\in \Gamma}\dist(z,
\sigma(H^{\#}_{\theta}))$,
the spectrum of $e^{-2\theta}H^{\#}_{}$ is at the distance $\ll
\inf_{z\in \Gamma}\dist(z, \sigma(H^{\#}_{\theta}))$ from the
spectrum of $H^{\#}_{\theta}$.
Using these estimates and using Eqn \eqref{ContourInt2},  we obtain
\begin{equation} \label{IX.5}
\|(H^{ }_{} - \lambda)^{-1} P_{>j} \| \le \frac{1}{\pi }
\oint_\Gamma [\dist(z,\sigma(H^{\#}_{\theta}))]^{-1} |z-
e^{-\theta}\lambda|^{-1}dz.
\end{equation}

We estimate the integrand on the r.h.s. of the above inequality. We
have for $\lambda \in Q_j$
$$|e^{\theta}z-\lambda|
\ge \sup_{ s \ge 0}(|e^{\theta}z +s-\lambda_j|-
|\lambda_j-s-\lambda|).$$ For $z \in \Gamma$, we have $ \inf_{ s \ge
0} |e^{\theta}z +s-\lambda_j|=
|z-\epsilon^{\#}_{j}|
=[(\frac{3}{4}(\epsilon^{\#}_{j+1}-\epsilon^{\#}_{j}))^2 +(\rIm
z)^2]^{1/2}$. Moreover, the the definition of $Q_j$ and \eqref{VI.2}
imply that $\sup_{\lambda \in Q_j}\inf_{ s \ge 0}
|\lambda_j-s-\lambda| \le \frac{1}{2}\delta_j \le
\frac{1}{2}\delta^{\#}_j$. Combining the last three estimates we
obtain
\begin{equation} \label{IX.6}
\inf_{\lambda \in Q_j}|e^{\theta}z-\lambda| \ge
\frac{1}{8}(\delta^{\#}_j + |\rIm z|).
\end{equation}

Next, we have for $z \in \Gamma$, $\dist(z, \sigma(H^{\#}_{\theta}))
=[(\epsilon^{\#}_{j+1} -(\frac{1}{4}\epsilon^{\#}_{j}+
\frac{3}{4}\epsilon^{\#}_{j+1}))^2 +(\rIm
z)^2]^{1/2}=[(\frac{1}{4}(\epsilon^{\#}_{j+1}-\epsilon^{\#}_{j}))^2
+(\rIm z)^2]^{1/2}, $ which gives
\begin{equation} \label{IX.7}
\dist(z, \sigma(H^{\#}_{\theta}))
 \ge
 \frac{1}{8}(\delta^{\#}_j +|\rIm z|).
\end{equation}
If $|\rIm \theta| \ll \delta^{\#}_j$, then estimates \eqref{IX.5} -
\eqref{IX.7} give
\begin{equation} \label{IX.8}\|(H^{}_{} -
\lambda)^{-1} P_{>j} \| \le C (\delta^{\#}_j)^{-1}.
\end{equation}
%
This together with the estimate \eqref{IX.3} yields
$$\|(H^{}_{} - \lambda)^{-1} \| \le C
(\delta^{\#}_j \sin(\rIm \theta))^{-1}.$$
This gives the desired estimate of the norm on the r.h.s. of
\eqref{kappa_j} for $\delta=0$.

Now we explain how to modify the above estimate in order to bound
the norm on the r.h.s. of \eqref{kappa_j} for $\delta>0$. First we
recall the definitions  $H_{}^{\delta} :=e^{-\varphi}H_{}
e^{\varphi}$ and $P_{j}^{\delta} := e^{-\varphi}P_{j}\ e^{\varphi}$
with $\varphi=\delta\langle x \rangle$. By a standard result, for
$\delta$ sufficiently small,
$$\sigma(H^{\delta})\bigcap \{\rRe z \le \nu\} = \sigma(H)\bigcap \{\rRe z \le
\nu\}.$$
This and the boundedness of $P_{j}^{\delta}$ show that the estimate
\eqref{IX.3} remains valid if we replace the operators $H$ and
$P_{<j}$ by the operators $H^{\delta}$ and $P^{\delta}_{<j}$.

Now to prove the estimate \eqref{IX.8} with the operator $H$
replaced by the operator $H^{\delta}$ we use in addition to the
estimates above the estimate $\big\| R^{\delta}(z)\big\| \le 2
\big\| R(z)\big\|
$ for $z \in \Gamma$ which is proven as follows.
%
%
%
By an explicit computation, $H_{}^{\delta} = H_{}+W $, where
$$W:=e^{\theta}(-\nabla\varphi \cdot \nabla -\nabla \cdot
\nabla\varphi -|\nabla\varphi|^2).$$ Hence for small $\delta$
(recall that $\varphi(x) := \delta \langle x \rangle$) the operator
$H_{}^{\delta}$ is a relatively small perturbation of the operator
$H_{}$. In particular, for $z \in \Gamma,\ \big\| R(z) W\big\| \le C
\delta \le 1/2$ and $R^{\varphi}(z) := [1- R(z)W]^{-1} R(z)$, where
$R(z) = (H_{pg} - z)^{-1} $ and $R^{\delta}(z) = (H_{pg}^{\delta}-
z)^{-1} $. Using the last two relations we estimate $\big\|
R^{\delta}(z)\big\| \le 2 \big\| R(z)\big\|
$ for $z \in \Gamma$. This, as was mentioned above, implies the
estimate \eqref{IX.8} with the operators $H$ and $P_{>j}$ replaced
by the operators $H^{\delta}$ and $P^{\delta}_{>j}$. This completes
the proof of the proposition.
%
%
%
%
%
%
%
%
%
%
\qed

Since $\epsilon^{\#}_{j}= \epsilon^{(p)}_j +O(g^2)$, we have that
$\delta^{\#}_{j} \ge \epsilon^{(p)}_{gap}(\nu)
- O(g^2)$ for $j \le j(\nu):= \max\{j: \epsilon^{(p)}_{j} \le
\nu\}$. Therefore the restriction $g \ll \min(\kappa^{}_{j},
\epsilon^{(p)}_{gap}(\nu))$, used above, is implied by the
restriction $$g \ll \epsilon^{(p)}_{gap}(\nu),$$ imposed in Theorem
\ref{thm-specHtheta}. As was mentioned in Section \ref{sec-III},
Theorem \ref{thm-specHtheta} and the Combes argument presented in
the paragraph containing Eqn \eqref{I.7} imply Theorems
\ref{thm-main} and \ref{thm-main2}, provided we choose $\theta$ to
be $g$-independent and satisfying $0< \rIm \theta \ll
\epsilon^{(p)}_{gap}(\nu)$.
%
%
%
%
%
%
%
%
This completes the proof of Theorem \ref{thm-main}.
%
\QED

%
%

\secct{Appendix A. The Smooth Feshbach-Schur Map} \label{sec-app}
%

In this appendix, we
%
%
describe properties of the \textit{isospectral smooth Feshbach-Schur
map} introduced in Section \ref{sec-V}.
%
%
%
%
%
In what follows $H_{g}= H_{0g} +I_{g} \in GH_\mu$ and we use the
definitions of Section \ref{sec-V}.

We define the following maps appearing in some identities below:
\begin{eqnarray} \label{Qchi}
Q_{\pi} (H_{g} - \lambda) & := & \pi \: - \: \bpi \,
(H_{\bpi}-\lambda)^{-1} \bpi I_{g} \pi \comma
\\  \label{Q2chi}
Q_{\pi} ^\#(H_{g} - \lambda) & := & \pi \: - \: \pi I_{g} \bpi \,
(H_{\bpi}-\lambda)^{-1} \bpi \period
\end{eqnarray}
Note that $Q_{\pi} (H_{g} - \lambda) \in \cB( \Ran\, \pi , \cH)$ and
$Q_{\pi}^\#(H_{g} - \lambda) \in \cB( \cH , \Ran\, \pi)$.

The following theorem, proven in \cite{BachChenFroehlichSigal2003}
(see \cite{GriesemerHasler} for some extensions), states that the
smooth Feshbach-Schur map of $H_{g} - \lambda$ is isospectral to
$H_{g} - \lambda$.
%
\begin{theorem} \label{thm-FeshSpec}
Let
$H_{g}= H_{0g} +I_{g}$ satisfy
\eqref{V.6}. Then, as was mentioned in Section \ref{sec-V}, the
smooth Feshbach-Schur map $F_{\pi}$ is defined on $H_{g} - \lambda$
and has the following properties:
\begin{itemize}
\item[(i)] $ \lambda \in \rho(H_{g} ) \Leftrightarrow 0 \in \rho(F_{\pi}
(H_{g} - \lambda))$, i.e. $H_{g} - \lambda$ is bounded invertible on
$\cH$ if and only if $F_{\pi} (H_{g} - \lambda)$ is bounded
invertible on $\Ran\, \chi$;
\item[(ii)] If $\psi \in \cH \setminus \{0\}$ solves $H_{g}\psi = \lambda \psi$
then $\vphi := \chi \psi \in \Ran\, \pi \setminus \{0\}$ solves
$F_{\chi} (H_{g} - \lambda) \, \vphi = 0$; \item[(iii)] If $\vphi
\in \Ran\, \chi \setminus \{0\}$ solves $F_{\pi} (H_{g} - \lambda)
\, \vphi = 0$ then $\psi := Q_{\pi} (H_{g} - \lambda) \vphi \in \cH
\setminus \{0\}$ solves $H_{g}\psi = \lambda \psi$; \item[(iv)] The
multiplicity of the spectral value $\{0\}$ is conserved in the sense
that $\dim \cern (H_{g} - \lambda) = \dim \cern F_{\pi} (H_{g} -
\lambda)$;
\item[(v)] If one of the inverses, $(H_{g} - \lambda)^{-1}$ or $F_{\tau,\pi} (H_{g} - \lambda)^{-1}$,
exists then so does the other and these inverses are related by
\begin{equation} \label{eq-II-6}
(H_{g} - \lambda)^{-1}  =  Q_{\pi} (H_{g} - \lambda) \: F_{\pi}
(H_{g} - \lambda)^{-1} \: Q_{\pi} (H_{g} - \lambda)^\# \; + \; \bpi
\, (H_{\bpi}- \lambda)^{-1} \bpi \comma
\end{equation}
and $$ F_{\pi} (H_{g}- \lambda)^{-1}  =  \pi \, (H_{g}-
\lambda)^{-1} \, \pi \; + \; \bpi \, (H_{0 g}- \lambda)^{-1} \bpi
\period$$
%
\end{itemize}
\end{theorem}
%


\secct{Appendix B.  Proof of Theorem ~\ref{thm-Hlambdaj}}
\label{sec-XI}

In this Appendix we prove Theorem~\ref{thm-Hlambdaj}. As was
mentioned in Section \ref{sec-VII}, the proof follows the lines of
the proof of Theorem IV.3
of \cite{FroehlichGriesemerSigal2008b} (cf. Theorem 3.8 of \cite{BachChenFroehlichSigal2003} and Theorem 28 of \cite{GriesemerHasler2}). It is similar to the proofs
of related results of
\cite{BachFroehlichSigal1998a,BachFroehlichSigal1998b}.
We begin with some preliminary results.
%
%

Recall the notation $H_g   \ = \   H_{0g} \; + \; I_g$ (see
\eqref{Hg}). According to the definition (Eqn \eqref{Fesh})
of the smooth Feshbach-Schur map, $F_{\pi}$, we have that
\begin{eqnarray} \label{eq-III-2-18}
\lefteqn{ F_{\pi} \big( H_g - \lambda \big) \ = \   H_{0g} -
\lambda\; + \; \pi I_g \pi}
\\[1mm] \nonumber & & \hspace{-3mm}
\; - \; \pi \, I_g \, \bar{\pi} \big(H_{0g}- \lambda + \, \bar{\pi}
I_g \bar{\pi} \big)^{-1} \bar{\pi} \, I_g \, \pi \period
\end{eqnarray}
%
%
%
Here, recall,  $\pi \equiv \pi[H_f]$ is defined in \eqref{pi} and
$\bar{\pi} \equiv \bar{\pi}[H_f ] := \mathbf{1} - \pi[H_f ]$.
Note that, due to Eqn \eqref{V.16},
%
%
%
%
the Neumann series expansion in $\bpi I_g \bar{\pi}$ of the
resolvent in (\ref{eq-III-2-18}) is norm convergent and yields
\begin{equation} \label{eq-III-2-20}
 F_{\pi} \big( H_g - \lambda \big) \ = \ H_{0g} - \lambda + \sum_{L=1}^\infty
(-1)^{L-1} \, \pi I_g \bigg(  (H_{0g}- \lambda)^{-1}\bar{\pi}^2 \;
I_g \bigg)^{L-1} \pi \period
\end{equation}
%


%

To write the Neumann series on the right side of (\ref{eq-III-2-20})
in the generalized normal form we use Wick's theorem,
which we formulate now.

We begin with some notation. Recall the definition of the spaces $
GH^{m n}_\mu$ in Section \ref{sec-IV}. For $W_{m,n} \in  GH^{m
n}_\mu$ of the form \eqref{Imn}, we denote $W_{m,n}\equiv
W_{m,n}[\uw]
$, where $\uw := (w_{m,n})_{1 \le m+n \le 2}$ with $w_{m,n}$
satisfying \eqref{Ibnd} (not to confuse with the definitions of
Section \ref{sec-VI}). We introduce the operator families
\begin{eqnarray} \label{VIII.4}
\lefteqn{ W_{p,q}^{m,n} \big[ \uw \big| \, k_{(m,n)} \big] \ :=
\;\int_{B_1^{p+q}} \frac{ dx_{(p,q)} }{ |x_{(p,q)}|^{1/2} } \; a^*(
x_{(p)} ) \, }
\\ \nonumber & & \hspace{-7mm}
\times  w_{m+p,n+q} \big[  \, k_{(m)}, x_{(p)} \, , \,
                       \tk_{(n)}, \tx_{(q)} \big] \, a( \tx_{(q)} ),
\end{eqnarray}
for $m+n \geq 0$ and a.e. $k_{(m,n)} \in B_1^{m+n}$. Here we use the
notation for $x_{(p,q)}$, $x_{(p)}$, $\tx_{(q)}$, etc. similar to
the one introduced in Eqs.~(\ref{III.2})--(\ref{III.4}). For $m=0$
and/or $n=0$, the variables $k_{(0)}$ and/or $\tk_{(0)}$ are dropped
out.
%
Denote by $S_m$ the group of permutations of $m$ elements. Define
the symmetrization operation as
\begin{eqnarray} \label{eq-III-2-24}
\lefteqn{ w_{m,n}^{(\sym)}[ \, k_{(m,n)} ] }
\\ \nonumber & := &
\frac{1}{m! \, n!} \sum_{\pi \in S_m} \sum_{\tpi \in S_n} w_{m,n} [
 \, k_{\pi(1)}, \ldots, k_{\pi(m)} \, ; \, \tk_{\tpi(1)}, \ldots,
\tk_{\tpi(n)} \, ] .
\end{eqnarray}
Finally, below we will use the notation
%
%
%
\begin{eqnarray}  \label{eq-III-1-6.2} && \Sigma[ k_{(m)} ]  :=  |k_1| + \ldots +
|k_m| ,\\ \label{eq-III-2-25.2} &&  k_{(M,N)} =
(k^{(1)}_{(m_1,n_1)}, \ldots, k^{(L)}_{(m_L,n_L)}) \comma
\hspace{5mm} k^{(\ell)}_{(m_\ell,n_\ell)} \ \; = \ \;
(k^{(\ell)}_{(m_\ell)}, \tk^{(\ell)}_{(n_\ell)}) \comma
\\ \label{eq-III-2-26}
&& r_\ell :=  \Sigma[\tk^{(1)}_{(n_1)}]  + \ldots +
\Sigma[\tk^{(\ell-1)}_{(n_{\ell-1})}] \, + \,
\Sigma[k^{(\ell+1)}_{(m_{\ell+1})}] + \ldots +
\Sigma[k^{(L)}_{(m_L)}] \comma \hspace{11mm}
\\  \label{eq-III-2-27}
&& \tr_\ell  :=  \Sigma[\tk^{(1)}_{(n_1)}]  + \ldots +
\Sigma[\tk^{(\ell)}_{(n_{\ell})}] \, + \,
\Sigma[k^{(\ell+1)}_{(m_{\ell+1})}]  + \ldots +
\Sigma[k^{(L)}_{(m_L)}],
\end{eqnarray}
with $r_\ell =0$ if $n_1 =\ldots n_{\ell-1} = m_{\ell+1} = \ldots
m_L =0$ and similarly for $\tr_\ell$ and $m_1 + \ldots + m_L = M,\
n_1 + \ldots + n_L = N$.
%
%
%
\begin{theorem}[Wick Ordering] \label{thm-III-2-3}
Let 
$W_{m,n} \in GH^{m n}_\mu,\ m+n \geq 1$ and $F_j\equiv F_j[H_f], j=
0 \ldots L,$ where $F_j[r]$ are operators on the particle space
which are $C^s$ functions of $r$ and satisfy the estimates
$\|\langle p \rangle^{-2+n}F_j[r]\langle p \rangle^{-n} \| \le C$
for $n=0,1,2$.
Write $W := \sum_{m+n \geq 1} W_{m,n}$ with $W_{m,n} :=
W_{m,n}[w_{m,n}]$. Then
\begin{equation} \label{eq-III-2-23}
F_0 \, W \, F_1 \, W \cdots W \, F_{L-1} \, W \, F_L \
 = \ P_{pj}\otimes\ \tW \comma
\end{equation}
where $\tW:=\tW[ \tuw ],\ \tuw := (\tw_{M,N}^{(\sym)})_{M+N \geq 0}
$ with $\tw_{M,N}^{(\sym)}$ given by the symmetrization w.~r.~t.\
$k_{(M)}$ and $\tk_{(N)}$, of the coupling functions
\begin{eqnarray}  \nonumber
\lefteqn{ \tw_{M,N}[ r ; \, k_{(M,N)} ]  =  \sum_{m_1 + \ldots + m_L
= M, \atop n_1 + \ldots + n_L = N} \sum_{p_1, q_1, \ldots, p_L, q_L:
\atop m_\ell + p_\ell + n_\ell + q_\ell \geq 1} \prod_{\ell = 1}^L
\bigg\{ {m_\ell + p_\ell \choose p_\ell} {n_\ell + q_\ell \choose
q_\ell} \bigg\}}
\\ \nonumber
\\ \nonumber
& \hspace{-6mm} & F_0[r+\tr_0] \, \bigg\la \psi^{(p)}_j\otimes\Om
\bigg| \, \tW_1 \big[  \, k^{(1)}_{(m_1, n_1)} \big] \;
F_1[\hf+r+\tr_1] \; \tW_2 \big[  \, k^{(2)}_{(m_2, n_2)} \big]
\\  \label{eq-III-2-25}
& \hspace{-6mm} &  \cdots F_{L-1}[\hf+r+\tr_{L-1}] \; \tW_L \big[
 \, k^{(L)}_{(m_L, n_L)} \big] \; \psi^{(p)}_j\otimes\Om
\bigg\ra F_L[r+\tr_L]\comma
\end{eqnarray}
with
%
\begin{eqnarray} \label{VIII.12}
\tW_\ell \big[  \, k_{(m_\ell, n_\ell)} \big] & := &
W_{p_\ell,q_\ell}^{m_\ell,n_\ell} [ \uw \big| \:  \, k_{(m_\ell,
n_\ell)} ].
\end{eqnarray}
\end{theorem}
The proof of this theorem mimics the proof of
\cite[Theorem~A.4]{BachFroehlichSigal1998b}.

Next, we mention some properties of the scaling transformation. It
is easy to check that $S_\rho (\hf) = \rho \hf$, and hence
\begin{equation} \label{eq-III-2-3}
S_\rho ( \chi_\rho) = \ \chi_1 \hspace{5mm} \mbox{and} \hspace{6mm}
\rho^{-1}S_\rho \big( \hf \big) \ = \  \hf \comma
\end{equation}
%
%
%
which means that the operator $\hf$ is a \emph{fixed point} of
$\rho^{-1} S_\rho$. Further note that $E \cdot \one$ \emph{is
expanded} under the scaling map, $\rho^{-1} S_\rho(E \cdot \one) =
\rho^{-1} E \cdot \one$, at a rate $\rho^{-1}$.
%
%
%
%
Furthermore,
\begin{equation} \label{eq-III-2-5}
\rho^{-1} S_\rho \big( W_{m,n}[\uw] \big) \ = \ W_{m,n} \big[
s_\rho(\uw) \big]
\end{equation}
%
where the map $s_\rho$
is defined by $s_\rho(\uw):=(s_\rho(w_{m,n}) )_{m+n \geq 0} $ and,
for all $(m,n) \in \NN_0^2$,
\begin{equation} \label{eq-III-2-6}
s_\rho(w_{m,n}) \big[  k_{(m,n)} \big] \ = \ \rho^{m+n - 1} \:
w_{m,n}\big[  \; \rho \, k_{(m,n)} \big] \period
\end{equation}

As a direct consequence of Theorem~\ref{thm-III-2-3} and
Eqs.~\eqref{Fesh}, (\ref{eq-III-2-5})--(\ref{eq-III-2-6}) and
(\ref{eq-III-2-20}), we have

%
\begin{theorem} \label{thm-III-2-4}
Let $\lambda\in Q_j$ so that $H_g -\lambda \in \dom(\cR_\rho)$. Then
$\cR_\rho(H_g -\lambda)\mid _{\Ran (P_{pj}\otimes\
\textbf{1})}-\rho_0^{-1}(\lambda_{j}-\lambda)$ $= H(\huw)$ where the
sequence $\huw$ is described as follows: $\huw =
(\hw_{M,N}^{(sym)})_{M+N \geq 0}$ with $\hw_{M,N}^{(sym)}$, the
symmetrization w.~r.~t.\ $k^{(M)}$ and $\tk^{(N)}$ (as in
Eq.~(\ref{eq-III-2-24})) of the kernels
\begin{eqnarray} \label{VIII.14}
\lefteqn{ \hw_{M,N}[ \, r ; \, k_{(M,N)} ] \ = \ \rho^{M+N-1}
\,\sum_{L=1}^\infty (-1)^{L-1} \: \times }
\\ \nonumber
& \hspace{-5mm} & \sum_{m_1 + \ldots + m_L = M, \atop n_1 + \ldots +
n_L = N} \sum_{p_1, q_1, \ldots, p_L, q_L:
      \atop m_\ell + p_\ell + n_\ell + q_\ell \geq 1}
\prod_{\ell = 1}^L \bigg\{ {m_\ell + p_\ell \choose p_\ell} {n_\ell
+ q_\ell \choose q_\ell} \bigg\} \;  V_{\umpnq} [ r ; k_{(M,N)} ],
\end{eqnarray}
for $M+N \geq 1$, and
\begin{eqnarray} \label{VIII.14a}
\hw_{0,0}[ \, r ] \ = r +\ \rho^{-1} \,\sum_{L=2}^\infty (-1)^{L-1}
%
\sum_{p_1, q_1, \ldots, p_L, q_L:
      \atop p_\ell +  q_\ell \geq 1}
\prod_{\ell = 1}^L
 \;  V_{\upq} [ r ],
\end{eqnarray}
%
%
for $M=N =0$. Here $\umpnq := (m_1, p_1, n_1, q_1,$ $\ldots,$ $m_L,
p_L, n_L, q_L) \in \NN_0^{4L}$, and
\begin{eqnarray} \label{VIII.15}
\lefteqn{ V_{\umpnq} [ r ; k_{(M,N)} ] \ := \langle
\psi^{(p)}_j\otimes\Om,\ g^L F_0[\hf+r] \,}
\\ \nonumber
& & \hspace{-6mm}  \; \times \prod_{\ell = 1}^L \Big\{ \tW_\ell
\big[
               \, \rho k^{(\ell)}_{(m_\ell, n_\ell)} \big]
\; F_\ell[\hf+r] \Big\}\psi^{(p)}_j\otimes\Om\rangle \; .
\end{eqnarray}
with $M := m_1 + \ldots + m_L$,  $N := n_1 + \ldots + n_L$,
$F_\ell[r] \ :=P_{pj}\otimes \chi_1[r+\tr_\ell], $ for $\ell = 0,
L,$  and
%
\begin{equation} \label{VIII.35}
F_\ell[r] \ := \bar{\pi}[\rho (r+\tr_\ell)]^2 \big( H_{pg}+\rho
(r+\tr_\ell) -\lambda \big)^{-1},
\end{equation}
for $\ell = 1, \ldots, L-1$.  Here
the notation introduced in Eqs.~\eqref{VIII.4}--\eqref{eq-III-2-27}
and \eqref{VIII.12} is used.
\end{theorem}
%

%


We remark that Theorem~\ref{thm-III-2-4} determines $\huw$
only as a sequence of integral kernels that define an operator in
$\cB[\cF]$. Now we show that $\huw \in \cW^{\mu,s}$, i.e. $\| \huw
\|_{\mu, s,\xi} < \infty$. In what follows we use the notation
introduced in Eqs.~\eqref{VIII.4}--\eqref{eq-III-2-27}
and \eqref{VIII.12}. To estimate $\huw$, we start with the following
preparatory lemma
\begin{lemma} \label{lem-IX.3}
Let $\lambda\in Q_j$. For fixed $L \in \NN$ and $\umpnq \in
\NN_0^{4L}$, we have $V_{\umpnq} \in \cW_{M,N}^{\mu,s}$ and
\begin{equation} \label{VIII.16}
\big\| V_{\umpnq} \|_{\mu,s}  \leq \rho^{\mu+1} L^s \Big(\frac{C
g}{\rho}\Big)^{L}\;  \prod_{\ell =1}^{L} \big\| w_{m_\ell+p_\ell,
n_\ell+q_\ell} \big\|_{\mu}^{(0)}.
\end{equation}
\end{lemma}
\proof First we derive the estimate \eqref{VIII.16} for $\mu=0$.
Recall that the operators $\tW_{\ell}$ might be unbounded. To begin
with, we estimate
\begin{equation} \label{VIII.38}
\big| V_{\umpnq} [ r ; k_{(M,N)} ] \big|  \leq
g^L \big\| F_0[\hf+r] \big\| \; \prod_{\ell = 1}^L A_\ell \comma
\end{equation}
where $A_\ell := \Big\| \tW_\ell \big[
 \, \rho k^{(\ell)}_{(m_\ell, n_\ell)}
\big]F_\ell[\hf+r] \Big\|.$ We use the resolvents and cut-off
functions hidden in the operators $F_\ell[\hf+r]$ in order to bound
the creation and annihilation operators whenever they are present in
$\tW_{\ell}$.

Recall that the operator $F_\ell[\hf+r]$ we estimate below depends
on $\lambda$, see \eqref{VIII.35}. Now, we claim that for $\lambda
\in Q_j$
\begin{equation} \label{VIII.39}
\big\|(|p|^2 +\rho H_f+1) F_\ell[\hf+r] \big\|  \le C \rho^{-1}
\end{equation}
for $\ell=1, ..., L-1$ and $\big\|(|p|^2 +H_f+1) F_L[\hf+r] \big\|
\le C$. The last estimate is obvious. To prove the first estimate we
use the inequality
\eqref{Vbnd}, in order to convert the operator $|p|^2$ into the
operator $H_{pg}$:
$$\big\|(|p|^2 +\rho H_f+1) F_\ell[\hf+r] \big\| $$ $$\le
2\big\|(H_{pg} + \rho (H_f+r+\tr_\ell)+3) F_\ell[\hf+r] \big\|.$$
Clearly, it suffices to consider $\lambda$ changing in sufficiently
large bounded set. The the above estimate gives
\begin{equation} \label{IX.6}\big\|(|p|^2 +\rho H_f+1) F_\ell[\hf+r] \big\| \le
C\big\|F_\ell[\hf+r] \big\| +C.
\end{equation} If the operator $F_\ell[\hf+r]$ inside the
operator norm on the r.h.s. is normal, as in the case of the ground
state analysis, then its norm can be estimated in terms of its
spectrum. For non-normal operators we proceed as follows.
Using that $ \bar{\pi}[H_f]:=P_{pj}\otimes\chi_{H_f\ge\rho}+
\bar{P}_{pj}\otimes \mathbf{1}$,  we write
$$F_\ell[H_f+r] \ := P_{pj}\otimes[\chi_{s \ge \rho}]^2 \big(
\lambda_j+s -\lambda \big)^{-1}$$
\begin{equation} \label{F_ell}
+ \bar{P}_{pj}\otimes \mathbf{1}\big( \bar{H}_{pg}+s -\lambda
\big)^{-1},
\end{equation}
where $s:=\rho (H_f+r+\tr_\ell)$, recall, $\bar{P}_{pj}:= \mathbf{1}
- P_{pj}$ and $\bar{H}_{pg}:=H_{pg}\bar{P_{p}}$. Now, since $\rRe
(\lambda_j -\lambda) \ge -\rho/2 \ge - s/2$ for $\lambda \in Q_j$
and $s \ge \rho$, we have that $\lambda_{j}+s -\lambda \ge \rho/2$
for the first term on the r.h.s.. For the second term on the r.h.s.,
we observe that by the spectral decomposition of the operator $s$ in
\eqref{F_ell} we have
\begin{equation} \label{}
\sup_{\lambda \in Q_j}\| (\bar{P}_{pj}\otimes \mathbf{1})\big(
\bar{H}_{pg}+s -\lambda \big)^{-1}\| \le \sup_{\lambda \in Q_j,\ \mu
\ge 0}\| \bar{P}_{pj}\big( \bar{H}_{pg}+\mu -\lambda
\big)^{-1}\|_{part}.
\end{equation}
Since $Q_j -[0,\infty) = Q_j$ and due to \eqref{kappa_j} we have
\begin{equation} \label{}
\sup_{\lambda \in Q_j} \| (\bar{P}_{pj}\otimes \mathbf{1})\big(
\bar{H}_{pg}+s -\lambda \big)^{-1}\| \le \sup_{\lambda \in Q_j}
\|\bar{P}_{pj}\big( \bar{H}_{pg} -\lambda \big)^{-1}\|_{part} \le
\kappa_j^{-1}.
\end{equation}
Since $\rho \le \kappa_j$, the last estimate, together with the
estimate of the first term on the r.h.s. of \eqref{F_ell} mentioned
above, yields $\big\| F_\ell[\hf+r]\big\| \le C \rho^{-1}$ for
$\ell=1, ..., L-1$.
This, due to \eqref{IX.6}, implies the estimate \eqref{VIII.39}.


Next, since $\tW_\ell \big[  \, \rho k^{(\ell)}_{(m_\ell, n_\ell)}
\big]$ contain products of $p_\ell + q_\ell \le m_\ell + p_\ell +
n_\ell + q_\ell \le 2$  creation and annihilation operators (see
\eqref{VIII.4} and \eqref{VIII.12} and the paragraph after
\eqref{Hg}), we have, by \eqref{Ibnd}, \eqref{V.17}
- \eqref{V.19} and similar estimates (cf. \eqref{VI.10}), that
\begin{equation} \label{VIII.41}
\Big\| \tW_\ell \big[  \, \rho k^{(\ell)}_{(m_\ell, n_\ell)}
\big]\langle p\rangle^{-(2-s_\ell)}(H_f+1)^{-s_\ell/2} \Big\| \le C
\|w_{m'_\ell, n'_\ell}\|_{0}^{(0)}
,\end{equation}
where $m'_\ell:= m_\ell + p_\ell$ and  $n'_\ell:= n_\ell + q_\ell$
and $s_\ell:= m'_\ell + n'_\ell$ (remember that $s_\ell  \le 2$).
Consequently,
\begin{equation} \label{VIII.42}
A_\ell \le C \rho^{-1+\delta_{\ell,L}} \|w_{m'_\ell,
n'_\ell}\|_{0}^{(0)}.
\end{equation}

Now, since $\big\| F_0[\hf+r] \big\|_\op \le 1$ we find from
\eqref{VIII.38} and \eqref{VIII.41} that
\begin{equation} \label{VIII.43}
\big| V_{\umpnq} [ r ; k_{(M,N)} ] \big|  \leq \rho \Big(\frac{C
g}{\rho}\Big)^{L}\; \prod_{\ell = 1}^L  \big\| w_{m_\ell+p_\ell,
n_\ell+q_\ell}
       \big\|_{0}^{(0)}
\end{equation}
and similarly for the $r-$
derivatives. This proves the isotropic, \eqref{VIII.16} with $\mu
=0$, bound on the functions $V_{\umpnq} [ r ; k_{(M,N)} ]$.
%
%
%

Now we prove the anisotropic, $\mu >0$, bound on $V_{\umpnq} [ r ;
k_{(M,N)} ]$. Let $\varphi(x):=\delta \langle x\rangle$ for $\delta$
sufficiently small. Define for $\ell =1, ..., L-1$
$$F_\ell^{\delta}[H_f+r] \
:=e^{-\varphi}F_\ell[H_f+r]e^{\varphi}$$ and
$$\tW_\ell^{\delta} \big[  \,
k^{(\ell)}_{(m_\ell, n_\ell)} \big]:=e^{-\varphi}\tW_\ell \big[ \,
k^{(\ell)}_{(m_\ell, n_\ell)} \big]e^{\varphi}.$$
Note that this transformation effects only the particle variables.
%

%
%
%
Exactly in the same way as we proved the bounds   \eqref{VIII.39},
with $\ell =1, ..., L-1$, one can show the following estimates
\begin{equation} \label{VIII.44}
\big\|(|p|^2 +\rho H_f+1) F_\ell^{\delta}[\hf+r] \big\|
\le C \rho^{-1},
\end{equation}
provided $\lambda \in Q_j$ and $\delta \le \delta_0$.

Now, expression \eqref{VIII.15} can be rewritten for any $j$ as
\begin{eqnarray*}
\lefteqn{ V_{\umpnq} [ r ; k_{(M,N)} ]  \ :=g^L \ F_0[\hf+r] \,
e^{\varphi}\; \times}
\\ \nonumber
& & \hspace{-6mm}   \prod_{\ell = 1}^{j-1} \Big\{ \tW_\ell^{\delta}
\big[ \, \rho k^{(\ell)}_{(m_\ell, n_\ell)} \big] \;
F_\ell^{\delta}[\hf+r] \Big\} \times\;
\\ \nonumber
& & \hspace{-6mm} e^{-\varphi}\tW_j \big[  \, \rho k^{(j)}_{(m_j,
n_j)} \big] \; F_j[\hf+r]\times
\\ \nonumber
& & \hspace{-6mm} \prod_{\ell = j+1}^L \Big\{ \tW_\ell \big[
               \, \rho k^{(\ell)}_{(m_\ell, n_\ell)} \big]
\; F_\ell[\hf+r] \Big\}.
\end{eqnarray*}
Since, by the definition, the operator  $F_0[\hf+r]$ contains the
projection, $P_p$, we conclude that the operator $F_0[\hf+r]
e^{\varphi}$ is bounded. Hence
we obtain for $j=1, ..., L$
\begin{equation} \label{VIII.46}
\big| V_{\umpnq} [ r ; k_{(M,N)} ] \big|  \leq
C g^L \tilde{A}_j^{\delta}\prod_{\ell = 1}^{j-1} A_\ell^{\delta} \;
\prod_{\ell = j+1}^L A_\ell \comma
\end{equation}
where $A_\ell^{\delta} := \Big\| \tW_\ell^{\delta} \big[
 \, \rho k^{(\ell)}_{(m_\ell, n_\ell)}
\big]F_\ell^{\delta}[\hf+r] \Big\|$  and $\tilde{A}_j^{\delta}:=
\Big\|e^{-\varphi} \tW_\ell \big[
 \, \rho k^{(\ell)}_{(m_\ell, n_\ell)}
\big]F_\ell[\hf+r] \Big\|$. Furthermore, since $\tW_\ell
^{\delta}\big[  \, \rho k^{(\ell)}_{(m_\ell, n_\ell)} \big]$ contain
products of $p_\ell + q_\ell \le 2$ creation and annihilation
operators (see \eqref{VIII.4} and \eqref{VIII.12}), we have, by
\eqref{Ibnd}, \eqref{V.17}- \eqref{V.19} and similar estimates (cf.
\eqref{VI.10}),  that
$$\Big\| \tW_\ell ^{\delta}\big[  \, \rho
k^{(\ell)}_{(m_\ell, n_\ell)} \big]\langle
p\rangle^{-(2-s_\ell)}(H_f+1)^{-s_\ell/2} \Big\| $$
\begin{equation} \label{VIII.47}
 \le C
\|w_{m'_\ell, n'_\ell}\|_{0}^{(0)}
\end{equation}
and
$$\Big\|e^{-\varphi} \tW_\ell \big[  \, \rho
k^{(\ell)}_{(m_\ell, n_\ell)} \big]\langle
p\rangle^{-(2-s_\ell)}(H_f+1)^{-s_\ell/2} \Big\|$$
\begin{equation} \label{VIII.48}
\le C |\rho k^{(\ell)}_{(m_\ell, n_\ell)}|^\mu \|w_{m'_\ell,
n'_\ell}\|_{\mu}^{(0)}
,
\end{equation}
where $m'_\ell:= m_\ell + p_\ell$ and  $n'_\ell:= n_\ell + q_\ell$
and $s_\ell:= m'_\ell + n'_\ell$. Consequently,
\begin{equation} \label{VIII.49}
A_\ell^{\delta} \le C \rho^{-1} \|w_{m'_\ell, n'_\ell}\|_{0}^{(0)}\
\mbox{and}\ \tilde{A}_j^{\delta}\le C \rho^{\mu-1} | k^{(j)}_{(m_j,
n_j)}|^\mu \|w_{m'_j, n'_j}\|_{\mu}^{(0)}.
\end{equation}

Putting the equations \eqref{VIII.46}, \eqref{VIII.49} and
\eqref{VIII.42} together we arrive at
$$\big| V_{\umpnq} [ r ; k_{(M,N)} ] \big|  \leq \rho^{\mu+1}
 \Big(\frac{C g}{\rho}\Big)^{L}\;| k^{(j)}_{(m_j, n_j)}|^\mu \times$$
\begin{equation} \label{VIII.50}
 \big\| w_{m_j+p_j, n_j+q_j}
\big\|_{\mu}^{(0)} \; \prod_{\ell \neq j}^{1,L} \big\|
w_{m_\ell+p_\ell, n_\ell+q_\ell}
       \big\|_{0}^{(0)}
\end{equation}
and similarly for the $r-$derivatives.
Since any $i$, $k_i$ is contained, as a $3-$dimensional component,
in $k^{(j)}_{(m_j, n_j)}$ for some $j$, we find \eqref{VIII.16}.
\QED

%

Proof of Theorem~\ref{thm-Hlambdaj}. As was mentioned above we
present here only the case $s=1$, which is needed in this paper.
Recall that we assume $\rho \le 1/2$ and  we choose $\xi = 1/4$.
First, we apply Lemma~\ref{lem-IX.3} to \eqref{VIII.14} and use that
${m+p \choose p} \leq 2^{m+p}$. This yields
\begin{eqnarray} \label{VIII.22}
\lefteqn{ \big\| \hw_{M,N} \big\|_{\mu,s} \ \leq \ \sum_{L=1}^\infty
 \, \rho^\mu \, L^s \, \Big( \frac{C g}{\rho} \Big)^L \, \big(2
\,\rho \big)^{M+N} \: }
\\ \nonumber
& & \hspace{-7mm} \times \sum_{m_1 + \ldots + m_L = M, \atop n_1 +
\ldots + n_L = N} \! \sum_{p_1, q_1, \ldots, p_L, q_L:
      \atop m_\ell + p_\ell + n_\ell + q_\ell \geq 1}
\prod_{\ell = 1}^L \bigg\{
2^{p_\ell+q_\ell} \big\| w_{m_\ell+p_\ell, n_\ell+q_\ell}
       \big\|_{\mu}^{(0)} \bigg\} \period
\end{eqnarray}
%
Using the definition \eqref{VI.17} and the inequality $2\rho \le 1$,
we derive the following bound for $\hat{\uw}_1:=(\hw_{M,N})_{M+N
\geq 1}$,
\begin{eqnarray} \nonumber
\lefteqn{ \big\| \hat{\uw}_1 \|_{\mu, s, \xi} \ := \ \sum_{M+N \geq
1} \xi^{-(M+N)} \, \big\| \hw_{M,N} \|_{\mu,s} \hspace{30mm} }
\\ \nonumber
& \hspace{-5mm} \leq & 2\,  \rho^{\mu+1} \sum_{L=1}^\infty L^s \,
\bigg( \frac{C g}{\rho} \bigg)^{L} \sum_{M+N \geq 1} \sum_{m_1 +
\ldots + m_L = M, \atop n_1 + \ldots + n_L = N} \sum_{p_1, q_1,
\ldots, p_L, q_L:
      \atop m_\ell + p_\ell + n_\ell + q_\ell \geq 1}
\\ \nonumber
& \hspace{-5mm} & \prod_{\ell = 1}^L \bigg\{
(2 \, \xi)^{p_\ell+q_\ell}\; \xi^{- ( m_\ell + p_\ell + n_\ell +
q_\ell )}\: \big\| w_{m_\ell+p_\ell, n_\ell+q_\ell}
       \big\|_{\mu}^{(0)} \bigg\}
\end{eqnarray}
\begin{eqnarray} \nonumber
\lefteqn{
\leq 2\, \rho^{\mu+1}  \sum_{L=1}^\infty L^s \, \bigg( \frac{C
g}{\rho} \bigg)^{L} \hspace{30mm} }
\\ \nonumber
& \hspace{-5mm} & \big\{ \sum_{m+n \geq 1} \big( \sum_{p=0}^m ( 2 \,
\xi )^{p} \big) \, \big( \sum_{q=0}^n ( 2 \, \xi )^{q} \big) \,
\xi^{- (m+n)}\: \| w_{m,n} \|_{\mu}^{(0)} \big\}^L.
\end{eqnarray}
Let $\big\| \uw_1 \|_{\mu, \xi}^{(0)} \ := \sum_{M+N \geq 1}
\xi^{-(m+n)} \, \big\| w_{m,n} \|_{\mu}^{(0)} $, where, recall,
$\uw_1 := (w_{m,n})_{m+n \geq 1}$ (we introduce this norm in order
to ease the comparison with the results of
\cite{BachChenFroehlichSigal2003}). Using the assumption $\xi
= 1/4$ and the estimate $\sum_{p=0}^m ( 2 \, \xi )^{p} \le
\sum_{p=0}^\infty \big( 2 \, \xi \big)^{p} \ =  \frac{1}{1 \: - \: 2
\, \xi }$,  we obtain
\begin{eqnarray}
\label{eq-III-3-18}
\big\| \hat{\uw}_1 \|_{\mu, s, \xi} \leq & 2\, \rho^{\mu+1}
\sum_{L=1}^\infty L^s \, B^L,
\end{eqnarray}
where
\begin{equation} \label{VII.38}
B:= \frac{C g }{\rho(1 \: - \: 2 \, \xi)^2} \, \big\| \uw_1
\big\|_{\mu, \xi}^{(0)} .
\end{equation}
%
%

Our assumption $g \ll \rho$ also insures that
%
$B \
\leq \ \frac{1}{2} .$
%
Thus the geometric series on the r.h.s. of (\ref{eq-III-3-18}) is
convergent. We obtain for $s=0, 1$
\begin{equation} \label{eq-III-3-21}
\sum_{L=1}^\infty L^s \, B^L \
 \leq \ 8 \, B \period
\end{equation}
Inserting (\ref{eq-III-3-21}) into (\ref{eq-III-3-18}), we see that
the r.h.s. of \eqref{eq-III-3-18} is bounded by $16 \, \rho^{\mu+1}
\, B $ which, remembering the definition of $B$ and the choice
$\xi=1/4$, gives
\begin{equation} \label{est-W}
\big\| \hat{\uw}_1 \|_{\mu, s,\xi} \leq 64 \, C g \, \rho^{\mu } \:
\big\| \uw_1 \big\|_{\mu, \xi}^{(0)} \period
\end{equation}
%

Next, we estimate $\hw_{0,0}$. We analyze the expression
\eqref{VIII.14a}.
%
%
%
%
%
%
%
%
Using estimate Eq.~(\ref{VIII.16}) with $\underline{m}=0,
\underline{n}=0$ (and consequently, $M=0, N=0$), we find
\begin{equation} \label{eq-III-3-24}
\rho^{- 1} \: \big\| V_{\upq} \|_{\mu,s} \ \leq \ L^s \, \rho^{\mu}
\big( \frac{C g}{\rho} \big)^{L} \, \prod_{\ell = 1}^L
\big\|w_{p_\ell, q_\ell}       \big\|_{\mu}^{(0)}. \
\end{equation}
In fact, examining the proof of Lemma \ref{lem-IX.3}  more carefully
we see that the following, slightly stronger estimate is true
\begin{equation} \label{eq-III-3-24a}
\rho^{- 1} \: \sup_{r \in I} \big|\partial_r^s V_{\upq} [ r ] \big|
\ \leq  L^s \, \rho^{\mu+s} \big( \frac{C g}{\rho} \big)^{L} \,
\prod_{\ell = 1}^L
\big\|
w_{p_\ell, q_\ell}       \big\|_{\mu}^{(0)}.
\end{equation}
Now, using \eqref{eq-III-3-24a},  we obtain
\begin{eqnarray} \nonumber
&& \: \rho^{-1} \sum_{L=2}^\infty \sum_{p_1, q_1, \ldots, p_L, q_L:
\atop p_\ell + q_\ell \geq 1} \sup_{r \in I} \big|
\partial_r^s V_{\upq} [ r ] \big|
\\ \nonumber
& \leq &  \:
\rho^{s+\mu}\, \sum_{L=2}^\infty L^s \, \Big( \frac{C g}{\rho}
\Big)^L \, \bigg\{ \sum_{p+q \geq 1} \big\|
w_{p,q} \big\|_{\mu}^{(0)} \bigg\}^L
\\ \nonumber
& \leq &  \:
\rho^{s+\mu}\, \sum_{L=2}^\infty L^s \,
D^L,
\end{eqnarray}
where $D := C g \xi \rho^{-1} \|
\partial_r^s\uw_1 \|_{\mu,0, \xi}$ with, recall, $\uw_1:=( w_{m,n} )_{m+n \geq 1}$. Now,  similarly
to (\ref{eq-III-3-21}), using that $\sum_{L=2}^\infty L^s D^L \leq
12 D^2$, for $D$ satisfying $D \leq
1/2$ (recall $g \ll \rho$),
we find  for $s= 0, 1$
$$\rho^{-1} \sum_{L=2}^\infty \sum_{p_1, q_1, \ldots, p_L, q_L: \atop
p_\ell + q_\ell \geq 1} \sup_{r \in I} \big|
\partial_r^s V_{\upq} [ r ] \big| $$
\begin{eqnarray}
 \label{eq-III-3-24b}
\leq  \:
12\rho^{s+\mu}\, \Big( \frac{C g \, \xi}{\rho} \, \big\| \uw_1
\big\|_{\mu, \xi}^{(0)} \Big)^2.
\end{eqnarray}
%

Next,
Eqns.~\eqref{VIII.14a} and \eqref{eq-III-3-24b} yield
%
%
\begin{equation} \label{est-E}
\big| \hw_{0,0}[ 0] \,
\big| \ \leq \
12 \rho^{\mu} \Big( \frac{C g \, \xi}{\rho} \, \big\| \uw_1
\big\|_{\mu, \xi}^{(0)} \Big)^2.
\end{equation}
%
%
We find furthermore that
\begin{eqnarray} \label{est-T'}
\sup_{r \in [0,\infty)} \big| \partial_r \hw_{0,0}[  \, r ] \, - \,
1 \big|
\leq
12 \rho^{\mu+1}\, \Big( \frac{C g \, \xi}{\rho} \, \big\| \uw_1
\big\|_{\mu, \xi}^{(0)} \Big)^2.
\end{eqnarray}

Now, recall that
%
$\big\| \uw_1 \big\|_{\mu, \xi}^{(0)}\le C $ and $\xi =
1/4$. Hence Eqns \eqref{est-E}, \eqref{est-T'}
and \eqref{est-W} give \eqref{Hlambdaj} with $s=1,\ \alpha = 12
\rho^{\mu}\, \Big( \frac{ C g }{\rho} \, \Big)^2$, $\beta= 12
\rho^{\mu+1}\, \Big( \frac{ C g}{\rho} \Big)^2$ and $\gamma = C \,
\rho^{\mu} g$.
This implies the statement of Theorem~\ref{thm-Hlambdaj}.
\QED

%
%
%

%

%

\secct{Supplement A: Background on the Fock space, etc}
\label{sec-SA}
%
Let $ \fh$ be either $ L^2 (\RR^3, \mathbb{C}, d^3 k)$ or  $ L^2
(\RR^3, \mathbb{C}^2, d^3 k)$. In the first case we consider $ \fh$
as the Hilbert space of one-particle states of a scalar Boson or a
phonon, and in the second case,  of a photon. The variable
$k\in\RR^3$ is the wave vector or momentum of the particle. (Recall
that throughout this paper, the velocity of light, $c$, and Planck's
constant, $\hbar$, are set equal to 1.) The Bosonic Fock space,
$\cF$, over $\fh$ is defined by
\begin{equation} \label{eq-I.10}
\cF \ := \ \bigoplus_{n=0}^{\infty} \cS_n \, \fh^{\otimes n}
\comma
\end{equation}
where $\cS_n$ is the orthogonal projection onto the subspace of
totally symmetric $n$-particle wave functions contained in the
$n$-fold tensor product $\fh^{\otimes n}$ of $\fh$; and $\cS_0
\fh^{\otimes 0} := \CC $. The vector $\Om:=1
\bigoplus_{n=1}^{\infty}0$ is called the \emph{vacuum vector} in
$\cF$. Vectors $\Psi\in \cF$ can be identified with sequences
$(\psi_n)^{\infty}_{n=0}$ of $n$-particle wave functions,  which are
totally symmetric in their $n$ arguments, and $\psi_0\in\CC$. In the
first case these functions are of the form, $\psi_n(k_1, \ldots,
k_n)$, while in the second case, of the form $\psi_n(k_1, \lambda_1,
\ldots, k_n, \lambda_n)$, where $\lambda_j \in \{-1, 1\}$ are the
polarization variables.

In what follows we present some key definitions in the first case
only limiting ourselves to remarks at the end of this appendix on
how these definitions have to be modified for the second case. The
scalar product of two vectors $\Psi$ and $\Phi$ is given by
\begin{equation} \label{eq-I.11}
\la \Psi \, , \; \Phi \ra \ := \ \sum_{n=0}^{\infty}  \int
\prod^n_{j=1} d^3k_j \; \overline{\psi_n (k_1, \ldots, k_n)} \:
\vphi_n (k_1, \ldots, k_n) \period
\end{equation}

Given a one particle dispersion relation $\om(k)$, the energy of a
configuration of $n$ \emph{non-interacting} field particles with
wave vectors $k_1, \ldots,k_n$ is given by $\sum^{n}_{j=1}
\om(k_j)$. We define the \emph{free-field Hamiltonian}, $\hf$,
giving the field dynamics, by
%
\begin{equation} \label{eq-I.17a}
(\hf \Psi)_n(k_1,\ldots,k_n) \ = \ \Big( \sum_{j=1}^n \om(k_j) \Big)
\: \psi_n (k_1, \ldots, k_n) ,
\end{equation}
for $n\ge1$ and $(\hf \Psi)_n =0$ for $n=0$. Here
$\Psi=(\psi_n)_{n=0}^{\infty}$ (to be sure that the r.h.s. makes
sense we can assume that $\psi_n=0$, except for finitely many $n$,
for which $\psi_n(k_1,\ldots,k_n)$ decrease rapidly at infinity).
Clearly that the operator  $\hf$ has the single eigenvalue  $0$ with
the eigenvector $\Om$ and the rest of the spectrum absolutely
continuous.

With each function $\vphi \in \fh$ one associates an
\emph{annihilation operator} $a(\vphi)$ defined as follows. For
$\Psi=(\psi_n)^{\infty}_{n=0}\in \cF$ with the property that
$\psi_n=0$, for all but finitely many $n$, the vector $a(\vphi)
\Psi$ is defined  by
\begin{equation} \label{eq-I.12}
(a(\vphi) \Psi)_n (k_1, \ldots, k_n) \ := \ \sqrt{n+1 \,} \, \int
d^3 k \; \overline{\vphi(k)} \: \psi_{n+1}(k, k_1, \ldots, k_n).
\end{equation}
These equations define a closable operator $a(\vphi)$ whose closure
is also denoted by $a(\vphi)$. Eqn \eqref{eq-I.12} implies the
relation
\begin{equation} \label{eq-I.13}
a(\vphi) \Om \ = \ 0 \period
\end{equation}
The creation operator $a^*(\vphi)$ is defined to be the adjoint of
$a(\vphi)$ with respect to the scalar product defined in
Eq.~(\ref{eq-I.11}). Since $a(\vphi)$ is anti-linear, and
$a^*(\vphi)$ is linear in $\vphi$, we write formally
\begin{equation} \label{eq-I.14}
a(\vphi) \ = \ \int d^3 k \; \overline{\vphi(k)} \, a(k) \comma
\hspace{8mm} a^*(\vphi) \ = \ \int d^3 k \; \vphi(k) \, a^*(k)
\comma
\end{equation}
where $a(k)$ and $a^*(k)$ are unbounded, operator-valued
distributions. The latter are well-known to obey the \emph{canonical
commutation relations} (CCR):
\begin{equation} \label{eq-I.15}
\big[ a^{\#}(k) \, , \, a^{\#}(k') \big] \ = \ 0 \comma
\hspace{8mm} \big[ a(k) \, , \, a^*(k') \big] \ = \ \delta^3
(k-k') \comma
\end{equation}
where $a^{\#}= a$ or $a^*$.

Now, using this one can rewrite the quantum Hamiltonian $\hf$ in
terms of the creation and annihilation operators, $a$ and $a^*$, as
\begin{equation} \label{Hfa}
\hf \ = \ \int d^3 k \; a^*(k)\; \om(k) \; a(k) \comma
\end{equation}
acting on the Fock space $ \cF$.

More generally, for any operator, $t$, on the one-particle space $
\fh$ we define the operator $T$ on the Fock space $\cF$ by the
following formal expression $T: = \int a^*(k) t a(k) dk$, where the
operator $t$ acts on the $k-$variable ($T$ is the second
quantization of $t$). The precise meaning of the latter expression
can obtained by using a basis $\{\phi_j\}$ in the space $ \fh$ to
rewrite it as $T: = \sum_{j} \int a^*(\phi_j) a(t^* \phi_j) dk$.

To modify the above definitions to the case of photons, one replaces
the variable $k$ by the pair $(k, \lambda)$ and adds to the
integrals in $k$ also the sums over $\lambda$. In particular, the
creation and annihilation operators have now two variables: $a_
\lambda^\#(k)\equiv a^\#(k, \lambda)$; they satisfy the commutation
relations
\begin{equation} \label{eq-I.15}
\big[ a_{\lambda}^{\#}(k) \, , \, a_{\lambda'}^{\#}(k') \big] \ = \
0 \comma \hspace{8mm} \big[ a_{\lambda}(k) \, , \,
a_{\lambda'}^*(k') \big] \ = \ \delta_{\lambda, \lambda'} \delta^3
(k-k') .
\end{equation}
One can also introduce the operator-valued transverse vector fields
by
$$a^\#(k):= \sum_{\lambda \in \{-1, 1\}} e_{\lambda}(k) a_{\lambda}^\#(k),$$
where $e_{\lambda}(k) \equiv e(k, \lambda)$ are polarization
vectors, i.e. orthonormal vectors in $\mathbb{R}^3$ satisfying $k
\cdot e_{\lambda}(k) =0$. Then in order to reinterpret the
expressions in this paper for the vector (photon) - case one either
adds the variable $\lambda$ as was mentioned above or replaces, in
appropriate places, the usual product of scalar functions or scalar
functions and scalar operators by the dot product of
vector-functions or vector-functions and operator valued
vector-functions.

\secct{Supplement B: Nelson model} \label{sect-SA}
%

In this supplement we describe the Nelson model describing the
interaction of electrons with quantized lattice vibrations. The
Hamiltonian of this model is
%
\begin{equation} \label{Hn}
H_g^{N} \ = \ H^{N}_0 \, + \, I_{g}^{N} \comma
\end{equation}
acting on the state space, $\cH=\cH_\at \, \otimes \, \cF$, where
now $\cF$ is the  Fock space for phonons, i.~e. spinless, massless
Bosons.
%
%
Here $g$ is a positive parameter - a coupling constant - which we
assume to be small, and
\begin{equation} \label{H_0}
H^{N}_0 \ = \ H^{N}_p \, + \, \hf \comma
\end{equation}
where $H^{N}_p=H_p$ and $\hf$ are given in \eqref{Hp} and
\eqref{Hf}, respectively,  but, in the last case, with the scalar
creation and annihilation operators, $a$ and $a^*$, and where the
interaction operator is $I_{g}^{N}:=g I$ with
%
\begin{equation} \label{XIII.3}
I \ :=
\int \frac{ \kappa(k) \: d^3 k}{ |k|^{1/2} } \: \big\{ e^{-i k x} \,
a^*(k) \: + \: e^{i k x} \, a(k) \big\}
\end{equation}
(we can also treat terms quadratic in $a$ and $a^*$ but for the sake
of exposition we leave such terms out). Here, $\kappa=\kappa(k)$ is
a real function with the property that
\begin{equation} \label{XIII.4}
| \kappa (k)| \le \const \min\{1, |k|^\mu\} \comma
\end{equation}
with $\mu >0$, and
\begin{equation} \label{XIII.5}
\int \frac{d^3 k}{|k| } \: |\kappa(k)|^{2}  \ < \ \infty.
\end{equation}
In the following, $\kappa$ is fixed and $g$ varies. It is easy to
see that the operator $I$ is symmetric and bounded relative to
$H_0$,
with the zero relative bound (see \cite{RSIV} for the corresponding
definitions). Thus $H_g^{N}$ is self-adjoint on the domain of $H_0$
for arbitrary $g$. Of course, for the Nelson model we can take an
arbitrary dimension $d\geq 1$ rather than the dimension $3$.

The complex deformation of the Nelson hamiltonian is defined as
(first for $\theta \in \mathbb{R}$)
\begin{equation}
H_{g \theta}^{N} := U_{ \theta} H_{g}^{SM} U_{ \theta}^{-1} \ .
\label{XII.3}
\end{equation}
Under Condition (DA), there is a Type-A (\cite{Kato}) family
$H^{N}_{g \theta}$ of operators analytic in the domain $ |\rIm
\theta| < \theta_0$, which is equal to \eqref{XII.3} for $\theta \in
\mathbb{R}$ and s.t. $H^{N
*}_{g \theta} = H^{N}_{g \overline{\theta}}$,
\begin{equation}
H^{N}_{g \theta}  =  U_{\rRe\theta}  H^{N}_{g  i\rIm\theta}
U_{\rRe\theta}^{-1}. \label{XII.4}
\end{equation}
Furthermore, $H^{N}_{g \theta}$ can be written as
\begin{equation}
H_{g \theta}^{N} = H_{p\theta}^{N}    \otimes \mathbf{1}_f +
e^{-\theta} \mathbf{1}_{p} \otimes H_f + I_{g \theta}^{N} \ ,
\label{XI.5}
\end{equation}
where $H_{p \theta}^{N} := U_{p \theta} H_{p}^{N} U_{p \theta}^{-1}$
and $I_{g \theta}^{N} := U_{ \theta} I_{g}^{N} U_{ \theta}^{-1}$.

%
%
In the Nelson model case one can weaken the restriction on the
parameter $\rho$ to $\rho \gg g^2$. One proceeds as follows. Assume
for the moment that the parameter $\lambda$ is real. Then the
operator $R_0$ is non-negative and, due to Eqn \eqref{V.10} and Eqn
\eqref{VI.10}, with $m+n \le 1$, and the fact that the operator $I$
is a sum of creation and annihilation operators, we have
\begin{equation} \label{V.18a}
\big\| R_0^{1/2}I_g R_0^{1/2} \big\| \ \leq \ C \rho^{-1/2}g \comma
\end{equation}
where $R_0^{1/2}:=(H_{0g} -\lambda)^{-1/2}\overline{\pi}$. Hence the
following series
$$\sum_{n=0}^{\infty} R_0^{1/2}\big(gR_0^{1/2}I_g R_0^{1/2}\big)^n R_0^{1/2}$$
is well defined, converges absolutely and is equal to
$\overline{\pi}(H_{ \overline{\pi}} -\lambda)^{-1}\overline{\pi}$.
Estimating this series gives the desired estimate  \eqref{Ibnd} in
the case of real $\lambda$. For complex $\lambda$ we proceed in the
same way replacing the factorization $R_0 = R_0^{1/2}R_0^{1/2}$, we
used, by the factorization $R_0 = |R_0|^{1/2}U |R_0|^{1/2}$, where
$|R_0|^{1/2}:=|H_{0g}-\lambda|^{-1/2}\bar{\pi}$ and $U$ is the
unitary operator $U:=(H_{0g} -\lambda)^{-1}|H_{0g}-\lambda|$.


\vspace{3mm}   \noindent {\bf Acknowledgements:}

It is a pleasure to thank J\"urg Fr\"ohlich and Marcel Griesemer for
many stimulating discussions and enjoyable collaboration and Elliott
Lieb for useful remarks. A part of this work was done while the
author was visiting ETH Z\"urich, ESI Vienna, and IAS, Princeton. He
is grateful to these institutions for hospitality.
 \vspace{3mm}


%


\begin{thebibliography}{10}

\bibitem{AFFS}W. Abou Salem,  J. Faupin, J. Fr\"ohlich,
I.M.Sigal. On theory of resonances in non-relativisitc QED, e-print,
arXiv, 2007.


\bibitem{Arai1999}
A. Arai. Mathematical analysis of a model in relativistic quantum
electrodynamics. Applications of renormalization group methods in
mathematical sciences (in Japanese) (Kyoto, 1999)


\bibitem{AraiHirokawa}
A. Arai,  Masao Hirokawa. Ground states of a general class of
quantum field Hamiltonians. {\em Rev. Math. Phys.} 12 (2000), no. 8,
1085--1135.


\bibitem{BachChenFroehlichSigal2003}
V.~Bach, T. Chen, J.~Fr{\"{o}}hlich, and I.~M. Sigal.
\newblock Smooth Feshbach map and operator-theoretic
renormalization group methods.
\newblock {\em Journal of Functional Analysis}, 203, 44-92, 2003.

\bibitem{BachChenFroehlichSigal2006}
V.~Bach, T. Chen, J.~Fr{\"{o}}hlich, and I.~M. Sigal.
\newblock Smooth Feshbach map and operator-theoretic
renormalization group methods
\newblock {\em Journal of Functional Analysis}, 203, 44-92, 2006.

\bibitem{BachFroehlichPizzo1}
V.~Bach, J.~Fr{\"{o}}hlich, and A.~Pizzo.
\newblock Infrared-Finite Algorithms in QED: The Groundstate of an Atom Interacting with the
Quantized Radiation Field.
\newblock {\em  Communications in Mathematical Physics}
264, Issue: 1, 145 - 165, 2006.


\bibitem{BachFroehlichPizzo2}
V.~Bach, J.~Fr{\"{o}}hlich, and A.~Pizzo. \newblock An
infrared-finite algorithm for Rayleigh scattering amplitudes, and
Bohr's frequency condition. \newblock {\em Comm. Math. Phys.} 274,
no. 2, 457--486, 2007.

\bibitem{BachFroehlichPizzo3}
V.~Bach, J.~Fr{\"{o}}hlich, and A.~Pizzo. \newblock Infrared-Finite
Algorithms in QED II. The Expansion of the Groundstate of An Atom
Interacting with the Quantized Radiation Field, mp\_arc.


\bibitem{BachFroehlichSigal1995}
V.~Bach, J.~Fr{\"{o}}hlich, and I.~M. Sigal.
\newblock Mathematical theory of non-relativistic matter and radiation.
\newblock {\em Letters~in Math.~Physics}, 34:183--201, 1995.


\bibitem{BachFroehlichSigal1998a}
V.~Bach, J.~Fr{\"{o}}hlich, and I.~M. Sigal.
\newblock Quantum electrodynamics of confined non-relativistic particles.
\newblock {\em Adv.~in Math.~}, 137:299--395, 1998.

\bibitem{BachFroehlichSigal1998b}
V.~Bach, J.~Fr{\"{o}}hlich, and I.~M. Sigal.
\newblock Renormalization group analysis of spectral problems in quantum field
  theory.
\newblock {\em Adv.~in Math.~}, 137:205--298, 1998.

\bibitem{BachFroehlichSigal1999}
V.~Bach, J.~Fr{\"{o}}hlich, and I.~M. Sigal.
\newblock Spectral analysis for systems of atoms and molecules coupled to the
  quantized radiation field.
\newblock {\em Commun.~Math.~Phys.}, 207(2):249--290, 1999.

\bibitem{BachFroehlichSigalSoffer1999}
V.~Bach, J.~Fr{\"{o}}hlich, I.~M. Sigal, and A.~Soffer.
\newblock Positive commutators and spectrum of {P}auli-{F}ierz {H}amiltonian of
  atoms and molecules.
\newblock {\em Commun.~Math.~Phys.}, 207(3):557--587, 1999.



\bibitem{Berger} M.~Berger.
\newblock {\em  Nonlinearity and functional analysis. Lectures on
nonlinear problems in mathematical analysis.} Pure and Applied
Mathematics. Academic Press,
New York-London, 1977.


\bibitem{CEH}
I. Catto, P. Exner, Ch. Hainzl.  \newblock Enhanced binding
revisited for a spinless particle in nonrelativistic QED.
\newblock {\em J. Math. Phys.} 45, no. 11, 4174--4185, 2004.

\bibitem{CH}
I. Catto, Ch. Hainzl. \newblock Self-energy of one electron in
non-relativistic QED. \newblock {\em J. Funct. Anal.} 207, no. 1,
68--110, 2004.

\bibitem{Chen2001}
T.~Chen.
\newblock Operator-theoretic infrared renormalization and construction of
dressed 1-particle states.
\newblock e-print, mp-arc 01-310, 2001.


\bibitem{ChenFrohlichPizzo1}
T. Chen, J. Fr\"ohlich, A. Pizzo. \newblock Infraparticle Scattering
states in non-relativistic QED: I. The Bloch-Nordsieck paradigm.
e-print, arXiv:0709.2493


\bibitem{ChenFrohlichPizzo2}
T. Chen, J. Fr\"ohlich, A. Pizzo. \newblock Infraparticle scattering
states in non-relativistic QED: II. Mass Shell Properties. e-print,
arXiv:0709.2812


\bibitem{Cohen-TannoudjiDupont-RocGrynberg1991}
C.~Cohen-Tannoudji, J.~Dupont-Roc, and G.~Grynberg.
\newblock {\em Photons and Atoms -- Introduction to Quantum Electrodynamics}.
\newblock John Wiley, New York, 1991.

\bibitem{Cohen-TannoudjiDupont-RocGrynberg1992}
C.~Cohen-Tannoudji, J.~Dupont-Roc, and G.~Grynberg.
\newblock {\em Atom-Photon Interactions -- Basic Processes and Applications}.
\newblock John Wiley, New York, 1992.


\bibitem{DerezinskiJaksic2001}
J.~Derezi\'nski and V.~Jak\v{s}i\'{c}.
\newblock Spectral theory of {P}auli-{F}ierz operators.
\newblock {\em J.~Funct. Anal.}, 180(2):243--327, 2001.

\bibitem{Faupin2007}
J.~Faupin.
\newblock Resonances of the confined hydrogenoid ion and the Dicke
effect in non-relativisitc quantum electrodynamics.
\newblock Preprint 2007.


\bibitem{Feshbach1958}
H.~Feshbach.
\newblock Unified theory of nuclear reactions.
\newblock {\em Ann.~Phys.}, 5:357--390, 1958.

\bibitem{FroehlichGriesemerSchlein1}
J.~Fr\"ohlich, M.~Griesemer and B.~Schlein.
\newblock Asymptotic electromagnetic fields in models of quantum-mechanical matter
interacting with the quantized radiation field.
\newblock {\em  Advances in
Mathematics} 164, Issue: 2, 349-398, 2001.

\bibitem{FroehlichGriesemerSchlein2}
J.~Fr\"ohlich, M.~Griesemer and B.~Schlein.
\newblock Asymptotic completeness for Rayleigh scattering.
\newblock {\em Ann. Henri Poincar�} 3, no. 1,
107--170, 2002.

\bibitem{FroehlichGriesemerSchlein3}
J.~Fr\"ohlich, M.~Griesemer and B.~Schlein.
\newblock Asymptotic completeness for Compton scattering.
\newblock {\em Comm. Math. Phys.} 252, no. 1-3,
415--476, 2004.

\bibitem{FroehlichGriesemerSigal2008a}
J.~Fr\"ohlich, M.~Griesemer and I.M.~Sigal.
\newblock Spectral theory for the standard model
of non-relativisitc QED. {\em Comm. Math. Phys.}, 283, no 3,
613-646, 2008.

\bibitem{FroehlichGriesemerSigal2008b}
J.~Fr\"ohlich, M.~Griesemer and I.M.~Sigal.
\newblock Spectral renormalization group analysis. e-print, ArXiv, 2008.

\bibitem{FroehlichGriesemerSigal2008c}
J.~Fr\"ohlich, M.~Griesemer and I.M.~Sigal.
\newblock Spectral renormalization group and limiting absorption principle
for the standard model of non-relativisitc QED. In
preparation.


\bibitem{GergescuGerardMoeller2}
V.~Gergescu, C.~G\'erard, and J.S.~M\o ller.
\newblock Spectral Theory of massless {P}auli-{F}ierz models.
\newblock {\em Commun. Math. Phys.}, 249:29--78, 2004.

\bibitem{Griesemer}
M. Griesemer.
\newblock Non-relativistic matter and quantized radiation. e-print, arXiv

\bibitem{GriesemerHasler}
M.~Griesemer and D.~Hasler.
\newblock On the smooth {F}eshbach-{S}chur map.
\newblock {\em J. Funct. Anal.}, 254(9):2329--2335, 2008.

\bibitem{GriesemerHasler2}
M.~Griesemer and D.~Hasler.
\newblock Analytic perturbation theory and renormalization analysis of matter
  coupled to quantized radiation.
\newblock arXiv:0801.4458.



\bibitem{GriesemerLiebLoss}
M. Griesemer, E.H. Lieb and M. Loss.
\newblock Ground states in non-relativistic quantum electrodynamics. \newblock {\em Invent. Math.}
 145, no. 3, 557--595, 2001.

\bibitem{GustafsonSigal}
S. Gustafson and I.M. Sigal.
\newblock {\em  Mathematical Concepts of Quantum Mechanics.} 2nd edition.
\newblock Springer 2006.

\bibitem{H}
Ch. Hainzl. \newblock One non-relativistic particle coupled to a
photon field. \newblock {\em Ann. Henri Poincar�.} 4, no. 2,
217--237, 2003.


\bibitem{HVV}
Ch.~Hainzl, V.~Vougalter, S.~Vugalter.
\newblock  Enhanced binding in non-relativistic QED.
\newblock {\em Commun. Math. Phys.}, 233, no. 1, 13--26, 2003.

\bibitem{HaslerHerbst2007}
D. Hasler and I. Herbst. \newblock  Absence of Ground States for a
Class of Translation Invariant Models of Non-relativistic QED. ArXiv

\bibitem{HaslerHerbstHuber2007}
D. Hasler, I. Herbst and M.Huber.
\newblock  On the lifetime of quasi-stationary states in
non-relativisitc QED. ArXiv:0709.3856.

\bibitem{HillePhillips}
E. Hille and R.S.Phillips.
\newblock  Functional Analalysis and Semi-groups. AMS 1957.


\bibitem{Hirokawa2}
M. Hirokawa. \newblock  Recent developments in mathematical methods
for models in non-relativistic quantum electrodynamics.
\newblock {\em  A garden of quanta,}  209--242, World Sci.
Publishing, River Edge, NJ, 2003.

\bibitem{Hiroshima1}
F. Hiroshima. \newblock  Ground states of a model in nonrelativistic
quantum electrodynamics. I. \newblock {\em J. Math. Phys.} 40
(1999), no. 12, 6209--6222.

\bibitem{Hiroshima2}
F. Hiroshima. \newblock  Ground states of a model in nonrelativistic
quantum electrodynamics. II. J. Math. Phys. 41 (2000), no. 2,
661--674.

\bibitem{Hiroshima3}
F. Hiroshima. \newblock  Ground states and spectrum of quantum
electrodynamics of nonrelativistic particles. \newblock {\em Trans.
Amer. Math.} Soc. 353 (2001), no. 11, 4497--4528 (electronic).

\bibitem{Hiroshima4}
F. Hiroshima. \newblock  Self-adjointness of the Pauli-Fierz
Hamiltonian for arbitrary values of coupling constants. \newblock
{\em Ann. Henri Poincar\'e} 3 (2002), no. 1, 171--201.

\bibitem{Hiroshima5}
F. Hiroshima. \newblock  Nonrelativistic QED at large momentum of
photons.\newblock {\em A garden of quanta},  167--196, World Sci.
Publishing, River Edge, NJ, 2003.

\bibitem{Hiroshima6}
F. Hiroshima. \newblock  Localization of the number of photons of
ground states in nonrelativistic QED. \newblock {\em Rev. Math.
Phys}. 15 (2003), no. 3, 271--312.

\bibitem{Hiroshima7}
F. Hiroshima. \newblock  Analysis of ground states of atoms
interacting with a quantized radiation field.  \newblock {\em Topics
in the theory of Schr\"odinger operators}, World Sci. Publishing,
River Edge, NJ, 2004, 145--272.

\bibitem{HiroshimaSpohn}
F. Hiroshima, and H. Spohn. \newblock  Ground state degeneracy of
the Pauli-Fierz Hamiltonian with spin. \newblock {\em Adv. Theor.
Math. Phys.} 5 (2001), no. 6, 1091--1104.

\bibitem{HislopSigal}
P.~Hislop and I.M. Sigal. \newblock {\em Introduction to Spectral
Theory. Applications to Schr\"odinger Operators}. Applied
Mathematical Sciences, 113. Springer-Verlag, New York, 1996.

\bibitem{Howland1975}
J.S. Howland.
\newblock The Livsic matrix in perturbation theory.
\newblock {\em J.~Math.~Anal.~Appl.}, 50:415--437, 1975.

\bibitem{HuebnerSpohn1}
M.~H\"ubner and H.~Spohn.
\newblock  Radiative decay: nonperturbative approaches. \newblock {\em Rev. Math. Phys.} 7, no. 3, 363--387, 1995.

\bibitem{Hunziker}
W.~Hunziker. \newblock  Resonances, metastable states and
exponential decay laws in perturbation theory. \newblock {\em  Comm.
Math. Phys.} 132, no. 1, 177--188, 1990.

\bibitem{HunzikerSigal}
W.~Hunziker and I.M.~Sigal.
\newblock The quantum $N$-body problem. \newblock {\em  J. Math. Phys.} 41, no. 6, 3448--3510, 2000.

\bibitem{Kato}
T.~Kato. \newblock {\em Perturbation Theory of Linear Operators},
volume 132 of {\em Grundlehren der mathematischen Wissenschaften}.
\newblock Springer-Verlag, 2 edition, 1976.

\bibitem{King} Ch. King. Resonant decay of a two state atom
interacting with a massless non-relativistic quantized scalar field.
Comm. Math. Phys. \textbf{165}(3), 569-594, (1994).

\bibitem{LiebLoss} E.H. Lieb and M.~Loss. Existence of atoms and molecules in non-relativistic
quantum electrodynamics, Adv. Theor. Math. Phys. {\bf 7}, 667-710
(2003).  arXiv math-ph/0307046.



\bibitem{Mueck}
M.~M\"uck. \newblock Construction of metastable states in quantum
electrodynamics. \newblock {\em Rev. Math. Phys.} 16, no. 1, 1--28,
2004.

\bibitem{Pizzo2003}
A.~Pizzo.
\newblock One-particle (improper) States in Nelson�s massless model.
\newblock {\em Annales
Henri Poincar�} 4, Issue: 3, 439 - 486, June, 2003.

\bibitem{Pizzo2005}
A.~Pizzo.
\newblock Scattering of an infraparticle: The one particle sector in
Nelson�s massless model.
\newblock {\em Annales Henri Poincar�} 6, Issue: 3, 553 - 606 ,
2005.




\bibitem{RSIV}
M.~Reed and B. Simon, \newblock {\em Methods of Modern Mathematical
Physics, IV, Ananlysis of Operators}. Academic Press, 1978.


\bibitem{Schur}
J. ~Schur.
\newblock \"Uber Potenzreihen die im Inneren des Einheitskreises beschr\"ankt sind.
\newblock {\em J. reine u. angewandte Mathematik}, 205--232, 1917.



\bibitem{Spohn}
H. Spohn. \newblock {\em Dynamics of Charged Particles and their
Radiation Field}, Cambridge University Press, Cambridge, 2004.

\end{thebibliography}

\end{document}